\documentclass[american,10pt,letterpaper]{article}
\usepackage[T1]{fontenc}
\usepackage{geometry}
\usepackage{babel}
\usepackage{prettyref}
\usepackage{mathtools}
\usepackage{amsmath}
\usepackage{amsthm}
\usepackage{amssymb}
\usepackage{graphicx}
\usepackage{setspace}
\usepackage[numbers]{natbib}
\usepackage[unicode=true,
 bookmarks=true,bookmarksnumbered=false,bookmarksopen=true,bookmarksopenlevel=2,
 breaklinks=false,pdfborder={0 0 0},pdfborderstyle={},backref=false,colorlinks=false]
 {hyperref}
\hypersetup{pdftitle={How the connectivity structure of neuronal networks influences responses to oscillatory stimuli},
 pdfauthor={Hannah Bos, Jannis Schuecker, Moritz Helias},
 unicode=true, bookmarks=true,bookmarksnumbered=false,bookmarksopen=true, breaklinks=false,pdfborder={0 0 0},backref=false,colorlinks=false}
\usepackage{breakurl}

\makeatletter
\usepackage{ifthen}
\newboolean{isarxiv}

\usepackage{changepage}

\usepackage[utf8]{inputenc}

\usepackage{textcomp,marvosym}

\usepackage{fixltx2e}

\usepackage{amsmath,amssymb}

\usepackage{cite}

\usepackage{nameref}

\usepackage[right]{lineno}

\usepackage{microtype}
\DisableLigatures[f]{encoding = *, family = * }

\usepackage{rotating}

\usepackage{setspace} 
\raggedright
\usepackage[aboveskip=3pt,labelfont=bf,labelsep=period,justification=raggedright,singlelinecheck=off]{caption}

\makeatletter
\renewcommand{\@biblabel}[1]{\quad#1.}
\makeatother

\date{}

\usepackage{epstopdf}

\usepackage{lastpage,fancyhdr,graphicx}
\usepackage{epstopdf}
\fancyheadoffset[L]{2.25in}
\fancyfootoffset[L]{2.25in}
\usepackage{tabularx} 
\usepackage{multirow}
\usepackage{colortbl}
\usepackage{dsfont}

\usepackage{prettyref}

\newrefformat{cap}{\hyperref[#1]{Fig~\ref{#1}}}
\newrefformat{fig}{\hyperref[#1]{Fig~\ref{#1}}}
\newrefformat{tab}{\hyperref[#1]{Table ~\ref{#1}}}
\newrefformat{sec}{\hyperref[#1]{Sec.~\ref{#1}}}
\newrefformat{sub}{\hyperref[#1]{Sec.~\ref{#1}}}
\newrefformat{eq}{\hyperref[#1]{Eq~(\ref{#1})}}

\usepackage{braket}

\usepackage[OT1]{fontenc}

\usepackage{etoolbox}
\makeatletter
\patchcmd{\@startsection}{\@ssect{#3}{#4}{#5}{#6}}{\@dblarg{\@sect{#1}{\@m}{#3}{#4}{#5}{#6}}}{}{\PackageError{fix-unnumbered-sections}{Unable to patch \string\@startsection; are you using a non-standard document class?}\@ehd}
\makeatother

\makeatother

\begin{document}
\global\long\def\ms{\,\mathrm{ms}}
\global\long\def\taus{\tau_{\mathrm{syn}}}
\global\long\def\taum{\tau_{m}}
\global\long\def\s{\,\mathrm{s}}
\global\long\def\Hz{\,\mathrm{Hz}}
\global\long\def\mV{\,\mathrm{mV}}
\global\long\def\taur{\tau_{r}}
\global\long\def\iFtr#1#2{\mathfrak{F}^{-1}[#1](#2)}
\global\long\def\E{\mathrm{E}}
\global\long\def\I{\mathrm{I}}
\global\long\def\DC{\mathrm{\mathrm{DC}}}

\let\oldnameref\nameref \renewcommand{\nameref}[1]{\textit{``\oldnameref{#1}''}}

\setboolean{isarxiv}{true}

\vspace*{0.35in}

\begin{flushleft}

{
\Large \textbf
\newline
{How the connectivity structure of neuronal networks influences responses to oscillatory stimuli} 
\addcontentsline{toc}{section}{Title}
}\newline
\\
Hannah Bos \textsuperscript{1,*},
Jannis Sch\"ucker \textsuperscript{1},
Moritz Helias \textsuperscript{1,2}
\\
\bigskip
{\bf 1} Institute of Neuroscience and Medicine (INM-6) and Institute for Advanced Simulation (IAS-6) and JARA BRAIN Institute I, J\"ulich Research Centre, 52425 J\"ulich, Germany\\
{\bf 2} Department of Physics, Faculty 1, RWTH Aachen University, 52074 Aachen, Germany\\
\bigskip



* \href{mailto:h.bos@fz-juelich.de}{h.bos@fz-juelich.de}

\end{flushleft}

\section*{Abstract }

Propagation of oscillatory signals through the cortex is shaped by
the connectivity structure of neuronal circuits. The coherence of
population activity at specific frequencies within and between cortical
areas has been linked to laminar connectivity patterns. This study
systematically investigates the network and stimulus properties that
shape network responses. The results show how input to a cortical
column model of the primary visual cortex excites dynamical modes
determined by the laminar pattern. Stimulating the inhibitory neurons
in the upper layer reproduces experimentally observed resonances at
$\gamma$ frequency whose origin can be traced back to two anatomical
sub-circuits. We develop this result systematically: Initially, we
highlight the effect of stimulus amplitude and filter properties of
the neurons on their response to oscillatory stimuli. Subsequently,
we analyze the amplification of oscillatory stimuli by the effective
network structure, which is mainly determined by the anatomical network
structure and the synaptic dynamics. We demonstrate that the amplification
of stimuli, as well as their visibility in different populations,
can be explained by specific network patterns. Inspired by experimental
results we ask whether the anatomical origin of oscillations can be
inferred by applying oscillatory stimuli. We find that different network
motifs can generate similar responses to oscillatory input, showing
that resonances in the network response cannot, straightforwardly,
be assigned to the motifs they emerge from. Applying the analysis
to a spiking model of a cortical column, we characterize how the dynamic
mode structure, which is induced by the laminar connectivity, processes
external input. In particular, we show that a stimulus applied to
specific populations typically elicits responses of several interacting
modes. The resulting network response is therefore composed of a multitude
of contributions and can therefore neither be assigned to a single
mode nor do the observed resonances necessarily coincide with the
intrinsic resonances of the circuit.

\section*{Author Summary }

Oscillations are ubiquitously generated within and propagated between
biological systems, for example in ecosystems, cell biology and the
brain. Recordings from neural signals often show oscillations which
are either generated within the recorded brain area or imposed on
it from other areas. It is an open question in neuroscience how the
underlying network structure influences the interaction between internally
generated and externally applied oscillations. This study systematically
analyzes how these two types of oscillations are reflected in the
spectra produced by neural networks and whether oscillatory input
can be utilized to uncover dynamically relevant sub-circuits of the
neuronal network. Previous work showed that structured neural circuits
yield dynamic modes which act as filters on internally generated noise
of the neuronal activity. Here we show how these filters act on external
input and how their superposition can yield resonances in the network
response, which cannot directly be linked to the sub-circuits generating
oscillations within the network. Simulation and theoretical analysis
of a column model of the primary visual cortex show that oscillatory
input applied to the inhibitory population in the upper layer elicits
a resonance at $\gamma$ frequency in their response, which can be
traced back to two distinct anatomical sub-circuits.

\ifthenelse{\boolean{isarxiv}}{}{\linenumbers}

\section*{Introduction }

Oscillations in the $\gamma$-frequency range ($30-80\Hz$) are observed
ubiquitously in recordings of brain activity, such as the local field
potential (LFP) \citep{Buzsaki04,Buzsaki12_203,Bartos07}. On the
single cell level, these oscillations can be generated from neurons
with a preferred frequency at which they transmit signals. This band-pass
filtering can arise from either single cell properties, for example
sub-threshold resonances of cells which are driven by fluctuations
\citep{Richardson03}, or from strongly driven cells whose input-output
relation exhibits a peak at their firing rate \citep{Lindner01_2934}.
On the network level, certain connectivity patterns have been hypothesized
to facilitate very slow as well as fast oscillations in the $\gamma$-range
\citep{beltramo2013_227}, namely the inter-neuron $\gamma$ (ING)
and pyramidal inter-neuron $\gamma$ (PING) motif (\citealp{Whittington2000},
reviewed in \citealp{Buzsaki12_203}). Although numerous theoretical
studies shed light on the emergent behavior of theses dynamical motifs
in isolation, few studies considered the effect of their embedment
in larger networks, such as the layered structure of the cortex \citep{Binzegger04,Thomson07_19}.
Similarly, the dynamical interaction of the network motif with the
surrounding network has been neglected when interpreting results of
experimental studies gathering evidence for the ING motif \citep{Cardin2009}
using oscillatory stimuli. In this study we analyze how the responses
to oscillatory stimuli are shaped by the network alone and therefore
only consider populations of neurons with non-resonant input-output
relations.

Neural response properties \citep{Destexhe2014,Altwegg-Boussac2014}
as well as the emergence of oscillations in the $\gamma$-range \citep{Fisahn98}
depend on the dynamical state of the network, which can be altered
by externally applied stimuli. It is still a matter of debate which
stimuli (natural or noise stimuli) elicit $\gamma$ oscillations \citep{Brunet2014,Hermes15,Hermes2015b}
and whether $\gamma$ oscillations of different frequencies and peak
shapes, elicited by these stimuli, are of the same anatomical origins
\citep{Ray2011,Hermes15}. Changes of the excitability of neurons,
that could be induced by stimuli, have theoretically been shown to
have a strong impact on the oscillations generated within the network
\citep{Brunel03a}. 

Probing the anatomical origin of network oscillations generated
in the cortex has become more feasible since the emergence of optogenetic
experiments \citep{Luo08,Zhang07,han09}, in which individual groups
of neurons can be stimulated selectively. Evidence for $\gamma$ oscillations
being generated by the interaction of inter-neurons alone has been
gathered by means of periodic light stimulation in optogenetically
altered mice \citep{Cardin2009}. A theoretical study \citep{Tchumachenko2014}
reproduces the experimental results by the analytical and numerical
treatment of a network composed of excitatory and inhibitory neurons.
The explanation requires gap junctions and a subthreshold resonance
of the inhibitory neurons. Using Hodgkin-Huxley-type model neurons,
Tiesinga \citep{Tiesinga12} showed that the results of Cardin et
al. can be reproduced by a PING mechanism if the excitatory cells
have an additional slow hyperpolarizing current. This result strengthened
the previous statement of the author \citep{Tiesinga09} that experimental
setups using oscillatory stimuli cannot distinguish between underlying
ING and PING mechanisms.

We can summarize the difficulties that arise in the interpretation
of these results with respect to the origin of the observed oscillations
by three main points. First, it is still under debate how strongly
external stimuli interfere with the dynamical state of the network.
Histed et al. \citep{Histed14} pointed out that weak light impulses
have a linear effect on the population responses of mice in vivo,
which they found to be sufficiently predictive for changes in behavior.
Second, mean-field theory of recurrent networks needs to be extended
to incorporate oscillatory stimuli \citep{Tiesinga09}. Third, the
dynamical interaction of the connection pattern generating the oscillation
with the surrounding network needs to be taken into account.

\textbf{}Describing oscillations that arise on the population level
from weakly synchronized neurons, Ledoux et al. \citep{Ledoux11_1}
investigate how external input shapes the dynamic transfer function,
which describes the response of a neuron to small rate perturbations.
However, they do not discuss the implications of this alteration for
the dynamical properties to the population rate spectra in high-dimensional
recurrently connected populations. Employing a similar framework,
Barbieri et al. \citep{Barbieri2014} showed by comparison to experimentally
measured spectra that describing an input signal as a perturbation
around the stationary state suffices to predict a considerable amount
of the variance of the LFP.

In this work, population dynamics of spiking neurons are reduced to
a rate-based description by a combined approach using mean-field theory
to determine the stationary rates and linear response theory for the
dynamical properties of the fluctuations. The reduction can therefore
be understood as a two-step procedure. In the first step the stationary
rate of the population is determined by evaluation of the nonlinear
stationary transfer function \citep{Siegert51,Ricciardi77,Fourcaud02},
which depends on the mean and variance of the input to the population
(also referred to as the working or operating point). All fluctuations
around the working point are considered linear in the second step
of the reduction, yielding the dynamic transfer function of the populations
\citep{Brunel00,Brunel99,Schuecker15_transferfunction}. It has been
shown that this level of reduction suffices to describe oscillations
in neural networks, that are visible on the population but not on
the single neuron level \citep{Bos16_1}. This reduction effectively
maps the dynamics of each population composed of numerous neurons
to a single noisy rate unit, which filters its input by a dynamic
transfer function. Oscillations are therefore described as filtered
noise (as found in \citep{Burns2011}) and the neural network is reduced
to coupled units, where the connections between the units shape the
correlation structure of the network. 

Keeping this reduction procedure in mind, we start from a rate based
description to illustrate the phenomena that arise when considering
oscillatory input to neural networks. In the first section, we use
a negatively self-coupled population to analysize how different types
of stimuli are reflected in different response measures. In particular,
we consider large versus small and filtered versus non-filtered stimuli
and their influence on absolute versus relative response spectra.
In the second section, we study the contribution of the connectivity
structure to the emergence and visibility of resonances in network
responses by analyzing three characteristic network motifs composed
of one excitatory and one inhibitory population each. Building on
the insights gathered from low-dimensional coupled rate circuits analyzed
in the first two sections, the third part is concerned with the analysis
of resonances evoked by oscillatory stimuli in a microcircuit model
based on primary sensory areas \citep{Potjans14_785} comprising millions
of spiking neurons.

We here show that phenomenological rate models with certain connectivity
patterns suffice to explain resonance in the $\gamma$ range in response
to oscillatory stimuli supplied to the inhibitory neurons, which is
not visible when stimulating the excitatory neurons. In addition we
demonstrate, that two different oscillation generating mechanisms,
one involving only the inhibitory and one involving both the inhibitory
and the excitatory neurons, generate similar resonances. In general
terms, we show that the responses of sub-circuits in isolation are
different than the responses of a system which embeds this sub-circuit.
The fact that a complex system cannot be understood by the analysis
of its parts in isolation, but only in its entirety has been pointed
out before \citep{Ashby56}. \textbf{}

\section*{Results}

\subsection*{Dynamic responses of a self-coupled inhibitory population\label{subsec:I-I-loop}}

In this section, we analyze how input is processed in a negatively
self-coupled dynamical rate unit that produces a rhythm in the $\gamma$-frequency
range. The model is inspired by a population of inhibitory leaky-integrate-and-fire
(LIF) neurons. We here contrast large versus small stimuli as well
as stimuli that affect the input current to the unit versus the output
rate of the unit. Changes induced by the input are considered in the
spectrum as well as in the power ratio (the spectrum normalized by
the spectrum without input).

\begin{figure}[ht]
\centering{}\includegraphics[scale=0.8]{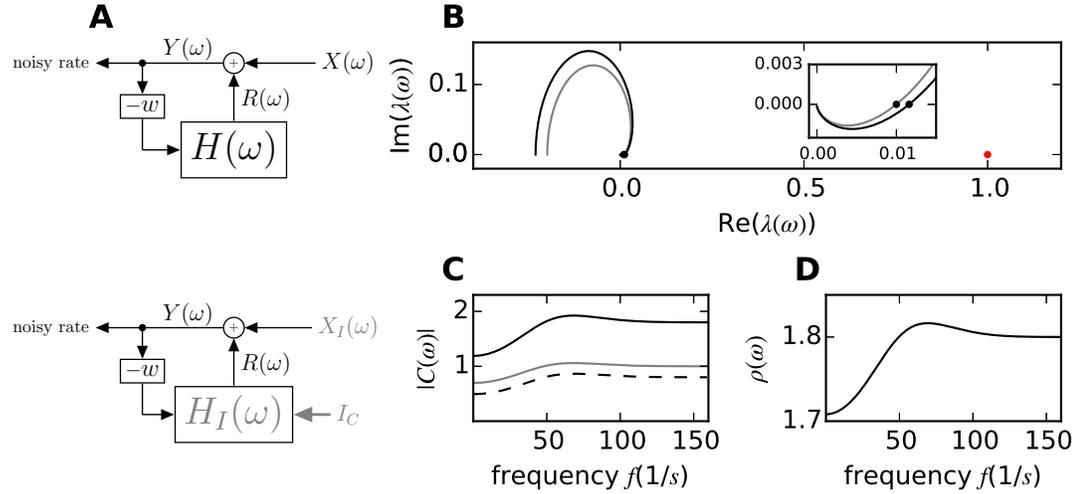}\caption{\textbf{Dynamics of a self-coupled inhibitory population with constant
stimulus. A }upper sketch: default circuit (\prettyref{eq:rate_self_consistent}).
The population characterized by the dynamic transfer function $H(\omega)$
emits the rate $R(\omega)$, which combined with the noise $X(\omega)$
yields the observed noisy rate $Y(\omega)$, which is amplified by
$-w$ and fed back to the population.\textbf{ }Lower sketch: circuit
as above\textbf{ }with additional constant input, which alters the
dynamic transfer function and the variance of the noise\textbf{. B}
Eigenvalue trajectories (Nyquist plot) $\lambda(\omega)=-wH(\omega)$
and $\lambda_{I}(\omega)=-wH_{I}(\omega)$ for the circuit with (black
curve) and without (gray curve) additional constant input, respectively.
The inset shows the values of the two trajectories for higher frequencies,
where the solid black dots denote their closest approach to the value
one. The close zoom shows that the eigenvalue trajectory assumes negative
imaginary values before it converges to zero. \textbf{C} Spectrum
of the circuit with ($C_{I}(\omega)$, black solid curve, \prettyref{eq:cross_spectrum})
and without ($C(\omega)$, gray solid curve, eq. \prettyref{eq:cross_spectrum_baseline})
additional constant input and their difference ($C_{I}(\omega)-C(\omega)$,
dashed black curve). \textbf{D} Power ratio of the spectra with and
without input $\rho(\omega)=C_{I}(\omega)/C(\omega)$. Parameters
of the circuit are specified in \prettyref{eq:params_1d_model}. \label{fig:eigenvalue_trajectory}}
\end{figure}

A sketch of the circuit with and without input is depicted in \prettyref{fig:eigenvalue_trajectory}A.
The dynamics of the circuit is determined by its dynamic transfer
function, which we here choose to approximate the dynamics in a corresponding
population of LIF neuron models with delays (for further detail see
\nameref{subsec:Static-and-dynamic}) as

\begin{equation}
H(\omega)=\frac{A\,e^{-i\omega d}}{1+i\omega\tau}\,e^{-\frac{\sigma_{d}^{2}\omega^{2}}{2}}.\label{eq:def_H}
\end{equation}

Here $\tau$ denotes the effective time constant of the dynamic transfer
function and $A$ its amplitude in response to a constant current.
The multiplicative factors $e^{-i\omega d}$ and $e^{-\frac{\sigma_{d}^{2}\omega^{2}}{2}}$
originate from the Gaussian distributed delays and with mean $d$
and variance $\sigma_{d}$. We here choose the variance equal to the
mean, i.e. $\sigma_{d}=d$. The first factor promotes oscillations,
while the second one suppresses the transfer of large frequencies.
In multi-dimensional systems, the generation of each peak in the spectrum
can be attributed to the dynamics of one eigenmode of the system,
where the dynamic transfer function of the $i$-th eigenmode is given
by the corresponding eigenvalue $\lambda_{i}(\omega)$ \citep{Bos16_1}.
Each mode emerges from the interplay of several populations. The transfer
function here could hence also be interpreted as the transfer function
of one dynamic mode. In this case its parameters are understood as
effective parameters, which are composed of the parameters of all
populations contributing to this mode.

The observed rate $Y$ of the unit is given by its output rate $R(\omega)$
combined with additive white noise $X(\omega)$ with zero mean $\langle X(\omega)\rangle=0$
and non-zero variance $\langle X(\omega)X^{\mathrm{T}}(-\omega)\rangle=D$
as

\begin{equation}
Y(\omega)=R(\omega)+X(\omega).\label{eq:rate_plus_noise}
\end{equation}
The noise term originates from the fact that the considered rate profile
actually describes a spike train. In other words, the spike train
can be considered as a noisy realization of the rate profile $R(\omega)$.
The internally generated noise in self-coupled populations of LIF
neurons exhibits a variance of $D=\frac{r_{0}}{N}$ \citep{Grytskyy13_131},
where $r_{0}$ denotes the stationary rate of the neurons in the population
and $N$ the number of neurons. The fluctuating rate produced by the
circuit (\prettyref{fig:eigenvalue_trajectory}A upper sketch) reads

\begin{align}
 & Y(\omega)=-wH(\omega)\,Y(\omega)+X(\omega)\nonumber \\
\Leftrightarrow & Y(\omega)=\frac{1}{1-\lambda(\omega)}X(\omega),\label{eq:rate_self_consistent}
\end{align}
where the fed back rate is weighted by the feedback strength $-w$
and we identify $\lambda(\omega)=-wH(\omega)$ as the eigenvalue of
the one-dimensional system. When considering LIF neuron models, the
strength $w$ is determined by the synaptic amplitude and the number
of connections. The spectrum of the population without additional
input is given by

\begin{equation}
C(\omega)=\langle Y(\omega)Y^{\mathrm{T}}(-\omega)=\left|\frac{1}{1-\lambda(\omega)}\right|^{2}D.\label{eq:cross_spectrum_baseline}
\end{equation}

\prettyref{fig:eigenvalue_trajectory}B shows the Nyquist plot of
the eigenvalue $\lambda(\omega)$, which determines the shape of the
spectrum. The peak frequency is determined by the point at which the
eigenvalue trajectory assumes its closest distance to unity, resulting
in a large prefactor in \prettyref{eq:cross_spectrum_baseline} (see
also \citep{Bos16_1}). The parameters of the dynamic transfer function
(\prettyref{eq:params_1d_model}) are based on the dynamic transfer
function of populations in a large scale model composed of LIF neurons
\citep{Potjans14_785} and chosen to produce a peak in the $\gamma$
frequency range (\prettyref{fig:eigenvalue_trajectory}C). The mapping
between the LIF neurons and rate models is described in the first
sections of the \nameref{sec:Methods}.

When considering the effect of external input to the spectrum in the
following, we distinguish weak and strong stimuli that require different
levels of description: Large input changes the stationary rate and
the dynamic properties of the population. Small input can be treated
as a perturbation around the stationary point which itself remains
unchanged. We will show in the last part of this study, that a small
oscillatory component in the input to a population is sufficient to
affect its spectrum considerably. We therefore neglect the effect
of the oscillatory component of the stimulus onto the stationary point
and restrict this analysis to either oscillatory input, which can
be treated as a perturbation, or oscillatory input with an additional
constant offset that may change the stationary state. 

We start by applying a large constant input to the population, yielding
an altered stationary rate of the system $r_{0}\rightarrow r_{0}+\delta r_{0}$.
This changes the spectrum for two reasons. First, the dynamic properties
of the population change yielding a new dynamic transfer function
$H(\omega)\rightarrow H_{I}(\omega)$ ($\lambda(\omega)\rightarrow\lambda_{I}(\omega)$).
This occurs in systems whose dynamic transfer function depends on
the statistics of the input, which is also referred to as the working
point. In general, an increase in the external rate can yield large
changes in the dynamic transfer function. However, reasonably sized
stimuli applied to populations in the fluctuation driven regime primarily
affect the offset of the transfer function and leave the shape approximately
unaltered (see \nameref{subsec:Approximation_of_dtf}). This suggests
the following approximation $H_{I}(\omega)\approx(1+\delta A/A)\,H(\omega)$
(see \prettyref{fig:eigenvalue_trajectory}B for the shifted eigenvalue
trajectory). Second, the input alters the stationary rate of the circuit
and therefore the amplitude of the internally generated noise $D\rightarrow D_{I}=(r_{0}+\delta r_{0})/N$.
The new spectrum is hence given by

\begin{equation}
C_{I}(\omega)=\left|\frac{1}{1-\lambda_{I}(\omega)}\right|^{2}D_{I}.\label{eq:cross_spectrum}
\end{equation}
In the following $C_{I}(\omega)$ is termed the response spectrum,
$\delta C(\omega)=C_{I}(\omega)-C(\omega)$ the excess spectrum, and
$\rho(\omega)=\frac{C_{I}\omega)}{C(\omega)}$ the power ratio. The
latter is commonly used in experimental studies since it is insensitive
to the filtering of the local field potential by the extracellular
tissue \citep{Bedard2006_118102} and dendritic morphology \citep{Linden11_859},
provided that both can be approximated as activity independent. All
three measures display a peak at the frequency generated by the circuit
(\prettyref{fig:eigenvalue_trajectory}C,D). The peak arises because
the excitatory input provided to the system effectively strengthens
the inhibitory loop that generates the oscillation by shifting the
eigenvalue closer to the value one and therefore closer to a rate
instability \citep{Bos16_1}. 

\begin{figure*}[htp]
\centering{}\includegraphics[scale=0.75]{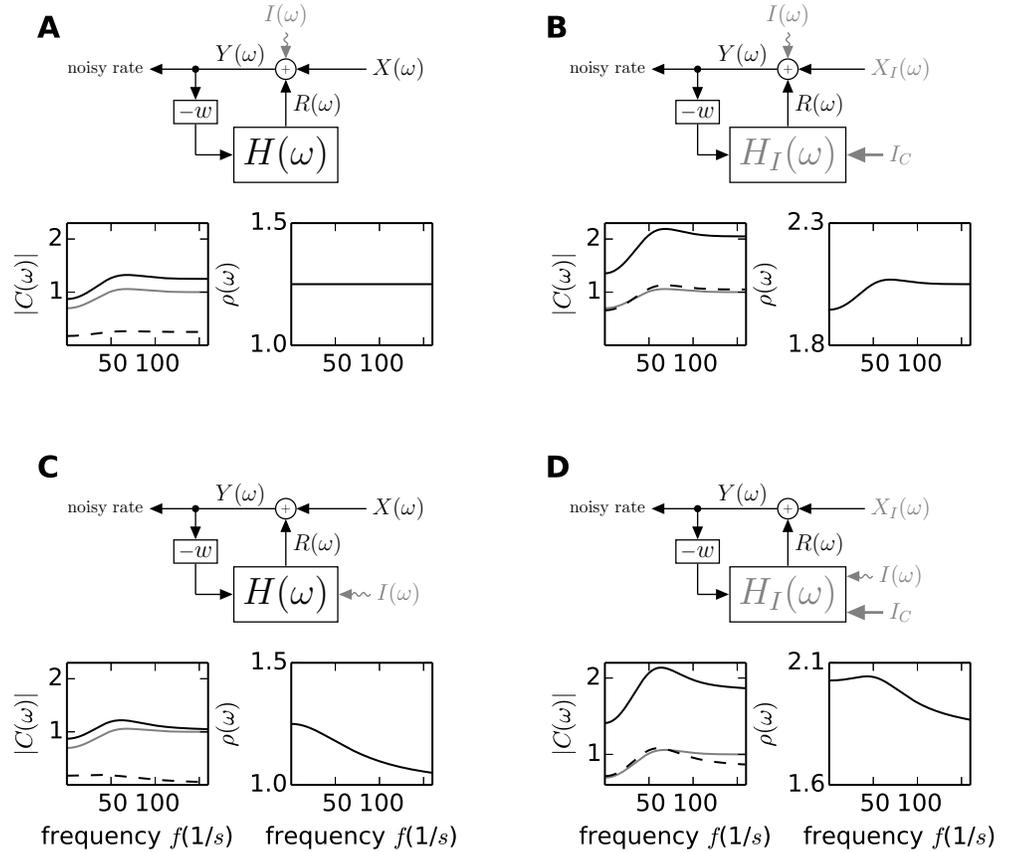}\caption{\textbf{Dynamics of a self-coupled inhibitory population with oscillatory
and constant stimuli.} The four panels depict four different types
of input for the circuit shown in \prettyref{fig:eigenvalue_trajectory}A.
In each panel the top shows a sketch of the circuit. The bottom left
panel shows the spectrum of the circuit at frequency $\omega$ without
input ($C(\omega)$, gray solid curve, \prettyref{eq:cross_spectrum_baseline})
and at stimulation frequency $\omega_{\protect\I}$ with input ($C_{I}(\omega)$,
black solid curve). Their difference ($C_{I}(\omega)-C(\omega)$)
as a function of the stimulation frequency $\omega_{\protect\I}$,
is depicted by the dashed black curve. The bottom left panel shows
the power ratio of the spectrum with and without input $\rho(\omega)=C_{I}(\omega)/C(\omega)$.
Parameters of the circuit are specified in \prettyref{eq:params_2d_model}.
\textbf{A }Circuit subject to a periodic modulation of the rate $\propto I(\omega)$
(\prettyref{eq:spec_stim}). The oscillatory input can be treated
as a perturbation and therefore added to the internally generated
noise. \textbf{B }Circuit subject to a periodic modulation of the
rate $\propto I(\omega)$ and a large constant component $I_{C}$
(\prettyref{eq:power_ratio_1d_alteredwp}). The latter changes the
dynamic transfer function and the variance of the noise. \textbf{C}
Input current modulated circuit (\prettyref{eq:spec_stim}). The oscillatory
input is send through the population filter before entering the circuit.
\textbf{D} Input current modulated circuit with a large constant component
(\prettyref{eq:power_ratio_1d_alteredwp}), that changes the dynamic
transfer function and the variance of the noise. \label{fig:fig1}}
\end{figure*}

Oscillatory input current injected into a neuron is necessarily filtered
by the dynamic transfer function of the neuron. It is, however, the
change in population firing rate that is recurrently processed on
the network level and eventually constitutes the measurable network
response. To this end we compare network responses to two types of
stimuli. The first type causes a modulation of the input current (``current
modulation'', CM), and the second type directly modulates the output
rate (``rate modulation'', RM, see also the illustrations in \prettyref{fig:fig1}B,C).
In the first case the stimulus is filtered by the dynamic transfer
function before it affects the activity of the population (see \nameref{subsec:spec_input}),
while it can be directly added to the rate in the latter case. The
spectrum of the stimulated network hence reads

\begin{eqnarray}
C_{I}(\omega) & = & \left|\frac{1}{1-\lambda(\omega)}\right|^{2}\Big(D+R_{\mathrm{I}}(\omega)R_{\mathrm{I}}^{*}(\omega)\Big)\nonumber \\
 &  & \mbox{with\,\,}R_{\mathrm{I}}(\omega)=\begin{cases}
I(\omega) & \mbox{for rate modulation (RM)}\\
H(\omega)I(\omega) & \mbox{for current modulation (CM).}
\end{cases}\label{eq:spec_stim}
\end{eqnarray}
Here $I(\omega)$ describes the stimulus in Fourier domain. Experimental
studies considered periodic stimuli to investigate the circuits underlying
the generation of oscillations \citep{Cardin2009}. Supplying a sinusoidal
stimulus with frequency $\omega_{I}$ ($I(\omega)=i\pi I_{0}\left(\delta(\omega+\omega_{I})-\delta(\omega-\omega_{I})\right)$)
to the one-dimensional circuit contributes an additional term to the
spectrum at stimulus frequency and yields the following power ratio

\begin{equation}
\rho(\omega)=\frac{C_{I}(\omega)}{C(\omega)}=1+\frac{1}{D}\begin{cases}
\pi^{2}I_{0}^{2}, & \mbox{\mbox{for rate modulation}}\\
\pi^{2}I_{0}^{2}\left|H(\omega)\right|^{2} & \mbox{for current modulation}
\end{cases}\:\mbox{and}\:\omega=\omega_{I}.\label{eq:power_ratio_inputnoise}
\end{equation}
This expression shows in particular, that the power ratio is independent
of the resonance properties of the circuit, since its contributions
(which are described in \prettyref{eq:cross_spectrum_baseline}) cancel.
The power ratio is independent of the stimulus frequency for rate
modulated systems (\prettyref{fig:fig1}A), while it reflects the
shape of the population filter $H$ in current modulated systems (\prettyref{fig:fig1}C).
This tendency is also reflected in the response spectrum, which displays
a constant offset in the RM system while for the CM system it approaches
the spectrum without stimulus for higher frequencies due to the low-pass
filter of the population.

This insight can be directly transferred to experimental studies.
To investigate the anatomical origin of oscillations, one seeks to
analyze the dynamics of the rate fluctuations generated within the
circuit. Hence, an upstream low-pass filter can give the false impression
that slow rate fluctuations are generated within the circuit, even
though they arise from the filter of the population. Adjusting the
input strength to emphasize fast frequencies can compensate for the
low-pass filtering of the populations; to this end one needs to replace
the amplitude as $I_{0}\rightarrow I_{0}/\left|H(\omega)\right|$.
An exact experimental implementation of this protocol is only possible,
if the dynamic transfer function of the stimulated neuronal population
is known. However, the relation can still be employed in an approximate
manner to counteract known influences. For example Tiesinga \citep{Tiesinga12}
modeled optogenetic stimuli as AMPA mediated currents. Since the dynamic
transfer function is a convolution of the synaptic and the population
filter, the synaptic filter could be counteracted in experiments by
considering the underlying receptor and neurotransmitter density,
which determine the time scales of the synaptic currents.

The two effects described above can be combined in a stimulus that
has a constant component, which increases the susceptibility and firing
rate of the target population, as well as an oscillatory component.
This yields the following power ratios at $\omega=\omega_{I}$

\begin{eqnarray}
\rho(\omega) & = & \left|\frac{1-\lambda(\omega)}{1-\lambda_{I}(\omega)}\right|^{2}\frac{D_{\I}+|I_{I}(\omega)|^{2}}{D}\nonumber \\
 & = & \left|\frac{1-\lambda(\omega)}{1-\lambda_{I}(\omega)}\right|^{2}\begin{cases}
\Big(\frac{D_{\I}}{D}+\frac{I_{0}^{2}}{D}\Big) & \mbox{\mbox{for rate modulation}}\\
\Big(\frac{D_{\I}}{D}+\frac{I_{0}^{2}\left|H_{I}(\omega)\right|^{2}}{D}\Big) & \mbox{for current modulation.}
\end{cases}\label{eq:power_ratio_1d_alteredwp}
\end{eqnarray}
From the analysis of the spectrum with constant input and from \prettyref{eq:cross_spectrum}
we know that the constant component of the stimulus shifts the eigenvalue,
such that the $\lambda$-dependent prefactor in the latter equation
displays a peak close to the internally generated frequency. This
peak, which reflects a positive change in the excitability of the
population ($D_{\I}>D\Rightarrow H_{I}(0),\lambda_{\I}(0)>H(0),\lambda(0)$),
is clearly visible in the rate modulated system (\prettyref{fig:fig1}B).
Here, the frequency independent contribution of the oscillatory stimulus
is added to the internal fluctuations. It therefore amplifies the
peak, which is shaped by the shift in the working point, but it cannot
affect the shape of the spectrum by itself. The spectrum for current-modulated
circuits experiences additional amplification at low frequencies compared
to the rate modulated circuit due to the multiplications of the dynamic
transfer function. If the change in excitability is large, this amplification
of low frequencies can overshadow the peak caused by the shift in
working point. The balance between these two effects depends on the
parameters of the population filter and the rise in excitability ($H_{I}(\omega)-H(\omega)$,
for small $\omega$) compared to the closeness of the system to an
instability, characterized by the term $\left|1-\lambda_{I}(\omega)\right|^{2}$(at
peak frequency).

In summary, the analysis of a one-dimensional self-coupled population
shows that the responses of the circuit can vary widely depending
on the properties of the system and the stimulus. Namely, a stimulus
that affects the stationary state of the system changes its dynamic
properties and therefore alters the strength of internally generated
oscillations. Stimuli that can be treated as perturbations amplify
the internally generated oscillation, but do not change the underlying
dynamical circuits that shape the spectrum. Input that directly affects
the rate of the population reveals information regarding the dynamics
of the circuit, while the responses to stimuli that are added to the
input current to the population run the risk of reflecting the filter
properties of the populations. These effects dominate the frequency-dependence
of power ratios, which become independent of the resonance properties
of the circuit. The response and the excess spectrum, in contrast,
both exhibit peaks at the internally generated oscillation. It is
therefore advantageous to consider one of the two latter quantities
in addition to the power ratio.

\subsection*{Stimulus evoked spectra in a two dimensional network\label{subsec:2d}}

In neural circuits, recurrent loops generating characteristic oscillations
do not appear in isolation, but are embedded into larger networks.
To analyze the effect of the surrounding network on the responses
elicited by oscillatory stimuli, we start by considering oscillation
generating circuits composed of one or two populations. We first describe
how the oscillations generated within the network can be understood
by means of dynamical modes and how the effect of a stimulus vector
can be split into components that each excite a different mode. The
responses of individual modes can in principal be traced back to an
anatomical circuit that generates the oscillation. However, the identification
of the origin of an oscillation is usually complicated by the fact
that responses to stimuli are composed of multiple modes. To isolate
this phenomenon we study three exemplary circuits in which evoked
responses can be treated as perturbations. Here, applied stimuli modulate
the rate directly (RM), assuming a stimulation protocol which counteracts
the filter properties of the population that receives the input. 

A two-dimensional circuit, as shown in \prettyref{fig:2d_ING1}A,
is composed of an excitatory (E) and an inhibitory (I) population,
with the dynamic transfer functions of population $i$ receiving input
from population $j$ 

\begin{equation}
H_{ij}(\omega)=\frac{A_{i}e^{-i\omega d_{j}}}{1+i\omega\tau_{i}}\,e^{-\frac{\sigma_{d_{j}}^{2}\omega^{2}}{2}},\:i,j\in{E,I},\label{eq:tf_2d}
\end{equation}
where $d_{j}$ denotes the delay of a connection starting at population
$j$. To illustrate the phenomena analyzed here in the simplest possible
setup, we assume that the neurons in the two populations have equal
working points and that the delays of all synapses are identically
distributed. The neurons therefore have equal stationary firing rates
$(r_{0,\E},r_{0,\I})=(r_{0},r_{0})$ as well as equal transfer functions
$H_{ij}(\omega)=H(\omega)$. The connectivity matrix is given by

\begin{equation}
\mathbf{W}=\left(\begin{array}{cc}
a & -b\\
c & -d
\end{array}\right)\label{eq:conn_matrix-1}
\end{equation}
with all parameters being positive $a,\:b,\:c,\:d>0$. In networks
of LIF-neuron models, these parameters are given by the product of
in-degrees and connection strength ($a=w_{EE}K_{EE}$, $b=w_{EI}K_{EI}$,
$c=w_{IE}K_{IE}$, and $d=w_{II}K_{II}$). The effective connectivity
matrix, which combines the anatomical and dynamic properties of the
circuit, is given by $\mathbf{M}(\omega)=H(\omega)\,\mathbf{W}$,
with the eigenvalues

\begin{equation}
\tilde{\lambda}_{0,1}(\omega)=H(\omega)\underset{=\lambda_{0,1}}{\underbrace{\left(\frac{a-d}{2}\pm\sqrt{\frac{(a-d)^{2}}{4}-(bc-da)}\right)}.}\label{eq:eigs_2d}
\end{equation}
The right ($\boldsymbol{u}_{i}$) and left ($\boldsymbol{v}_{i}$)
eigenvectors are given by

\begin{equation}
\mathbf{u}_{0}=\mathbf{v}_{0}=\left(\begin{array}{c}
1\\
0
\end{array}\right),\:\mathbf{u}_{1}=\mathbf{v}_{1}=\left(\begin{array}{c}
0\\
1
\end{array}\right)\quad\mbox{if}\:b=c=0\label{eq:evecs_iso}
\end{equation}
or 
\begin{align}
 & \mathbf{u}_{i}=\frac{1}{y_{i}}\left(\begin{array}{c}
1\\
x_{i}
\end{array}\right),\:\mathbf{v}_{0}=\frac{y_{0}}{x_{1}-x_{0}}\left(\begin{array}{c}
x_{1}\\
-1
\end{array}\right),\:\mathbf{v}_{1}=\frac{y_{1}}{x_{1}-x_{0}}\left(\begin{array}{c}
-x_{0}\\
1
\end{array}\right)\nonumber \\
 & \mbox{with}\:x_{i}=\begin{cases}
\frac{a-\lambda_{i}}{b} & \mbox{if}\:c=0\\
\frac{c}{\lambda_{i}+d} & \mbox{else}
\end{cases}\,\mbox{and}\,y_{i}=\sqrt{1+|x_{i}|^{2}}.\label{eq:evecs_net}
\end{align}

They are bi-orthogonal and normalized such that $|\mathbf{u}_{i}|^{2}=1$
and $\mathbf{v}_{i}^{\mathrm{T}}\mathbf{u}_{j}=\delta_{ij}$. Note
that in a more general setting, where the populations have different
stationary activities and therefore different transfer functions,
the eigenvectors are frequency dependent. The spectrum produced by
the circuit can be expressed as the sum of spectra produced by the
eigenmodes due to their auto- ($n=m$) and their crosscorrelation
($n\neq m$) (see \nameref{subsec:Composition-spec})

\begin{equation}
\mathbf{C}(\omega)=\sum_{n,m\in\{0,1\}}\frac{\beta_{nm}}{(1-\tilde{\lambda}_{n}(\omega))(1-\tilde{\lambda}_{m}^{*}(\omega))}\mathbf{u}_{n}\mathbf{u}_{m}^{\mathrm{T}*}\label{eq:spec_2d}
\end{equation}
with $\beta_{nm}=\sum_{i}D_{ii}\alpha_{n}^{i}\alpha_{m}^{i}$, where
$\alpha_{j}^{i}=\mathbf{v}_{j}^{\mathrm{T}}\mathbf{e}_{i}$ denotes
the projection of the $j$-th left eigenvector onto the $i$-th unit
vector with $\mathbf{e}_{i}^{\mathrm{T}}\in\left\{ (\begin{array}{cc}
1, & 0\end{array}),(\begin{array}{cc}
0, & 1\end{array})\right\} $. For LIF neuron models the diagonal elements of the stationary activity
matrix are given by $D_{ii}=r_{0,i}/N_{i}$. When describing neuronal
populations, the auto- and crosscorrelations of the modes describe
properties of groups of neurons, namely the summed correlations on
the single neuron level. For example, the autocorrelation of one modes
refers to the sum of all auto- and crosscorrelations between neurons
that constitute that mode.

The diagonal elements of $\mathbf{C}(\omega)$ (\prettyref{eq:spec_2d})
describe the spectra of the population activity. For frequencies $\omega_{c}$
at which one eigenvalue $\tilde{\lambda}_{c}(\omega_{c})$ approaches
the value one, a peak is visible in the spectrum. The anatomical connections
that determine the amplitude and frequency of this peak can be established
using the following quantities \citep{Bos16_1}

\begin{equation}
Z_{ij}^{\mathrm{amp}}=\left(\Re(Z_{ij}),\Im(Z_{ij})\right)\mathbf{k}^{\mathrm{T}}\quad\mbox{and}\quad Z_{ij}^{\mathrm{freq}}=\left(\Re(Z_{ij}),\Im(Z_{ij})\right)\mathbf{k}_{\perp}^{\mathrm{T}},\label{eq:def_sensitivity_measure}
\end{equation}
which identify the sensitivity of the eigenvalue to the connections
(defined by the matrix elements $M_{ij}$) via

\begin{equation}
Z_{kl}=\frac{v_{c,k}M_{kl}u_{c,l}}{\mathbf{v}_{c}^{\mathrm{T}}\mathbf{u}_{c}},\label{eq:def_Z}
\end{equation}
where $\mathbf{u}_{c}$ and $\mathbf{v}_{c}$ are the right and left
eigenvector associated to $\lambda_{c}(\omega_{c})$. The unit vectors
that describe directions in the complex plane: $\mathbf{k}$ points
from $\lambda_{c}(\omega)$ to the one and $\mathbf{\mathbf{k}_{\perp}}$
perpendicular to $\mathbf{\mathbf{k}}$, are given by

\begin{align}
\mathbf{k} & =(1-\Re(\lambda_{c}),\Im(\lambda_{c}))/\sqrt{(1-\Re(\lambda_{c}))^{2}+\Im(\lambda_{c})^{2}}\nonumber \\
\mathbf{k}_{\perp} & =(-\Im(\lambda_{c}),1-\Re(\lambda_{c}))/\sqrt{(1-\Re(\lambda_{c}))^{2}+\Im(\lambda_{c})^{2}},\label{eq:def_k_kper}
\end{align}
 where all dependencies on frequency were omitted for brevity of notation.

In electrophysiological recordings, the activities of the excitatory
and the inhibitory population are often indirectly observed via the
local field potential (LFP), which has been related to the input of
pyramidal neurons \citep{Mazzoni2015} (see \nameref{subsec:LFP})

\begin{equation}
C_{\mathrm{LFP}}(\omega)=\underset{:=C_{\mathrm{LFP}}^{\mathrm{auto}}(\omega)}{\underbrace{a^{2}C_{\E\E}(\omega)+b^{2}C_{\I\I}(\omega)}}+\underset{:=C_{\mathrm{LFP}}^{\mathrm{cross}}(\omega)}{\underbrace{2ab\Re\left(C_{\E I}(\omega)\right)}}.\label{eq:def_LFP}
\end{equation}
The LFP gets contributions from the excitatory and the inhibitory
current onto the excitatory neurons as well as from their crosscorrelation.
Defining the LFP in this way implicitly assumes $\delta$-shaped synaptic
currents, otherwise the contributions above would additionally be
filtered by the synaptic kernels (see \prettyref{eq:def_CLFP_method}).
Stimulating the circuit with sinusoidal input $F(\omega)=i\pi I_{0}\left[\delta(\omega+\omega_{I})-\delta(\omega-\omega_{I})\right]$
of frequency $\omega_{I}$ and amplitude $I_{0}$ in the direction
$\mathbf{e}_{I}$ elicits the excess spectrum (see \prettyref{eq:spectrum_decomp_with_input})
at $\omega=\omega_{I}$

\begin{equation}
\delta\mathbf{C}(\omega)=\mathbf{C}_{I}(\omega)-\mathbf{C}(\omega)=\sum_{n,m\in\{0,1\}}\underset{=\delta\mathbf{C}^{nm}(\omega)}{\underbrace{\frac{\beta_{nm}^{I}}{(1-\tilde{\lambda}_{n}(\omega))(1-\tilde{\lambda}_{m}^{*}(\omega))}\mathbf{u}_{n}\mathbf{u}_{m}^{\mathrm{T}*}}}\label{eq:excess_spec_2d}
\end{equation}
with $\beta_{nm}^{I}=\pi^{2}I_{0}^{2}\gamma_{n}\gamma_{m}^{*}$,
where $\gamma_{j}=\mathbf{v}_{j}^{\mathrm{T}}\mathbf{e}_{I}$ marks
the projection of the stimulus direction onto the left eigenvector.
If the stimulus vector is parallel to the right eigenvector of one
mode $\mathbf{e}_{I}=x\mathbf{u}_{k}$ it will excite only this mode
($\gamma_{k}=x,\,\gamma_{i\neq k}=0$). The scalar $\gamma_{i}$ therefore
measures the portion by which the $i$-th eigenmode is stimulated
by the stimulus vector $\mathbf{e}_{I}$. The stimulus-induced
component of the LFP response that originates in the autocorrelation
of the currents is given by

\begin{equation}
\delta C_{\mathrm{LFP}}^{\mathrm{auto}}(\omega)=C_{\mathrm{LFP},I}^{\mathrm{auto}}(\omega)-C_{\mathrm{LFP}}^{\mathrm{auto}}(\omega)=\sum_{n,m\in\{0,1\}}\left(a^{2}\delta C_{\E\E}^{nm}(\omega)+b^{2}\delta C_{\I\I}^{nm}(\omega)\right).\label{eq:def_dC_LFP_auto}
\end{equation}
Since the contributions of individual modes are more straightforward
to separate for the autocorrelations, we only consider the LFP contribution
defined in \prettyref{eq:def_dC_LFP_auto} in detail (see \prettyref{eq:def_CLFP_method}
for the full LFP spectrum). It will, however, be shown, that the crosscorrelations
do not interfere with the discussed effects.

In the following sub-sections, we will show on three exemplary circuits,
that non-negligible connectivity between populations evokes responses
of several modes when individual populations are stimulated. The interference
of these mode responses can yield similar network responses for different
underlying network structures.

\subsubsection*{A self-coupled inhibitory circuit embedded in a two-dimensional network\label{subsec:ING1}}

\begin{figure}[htp]
\centering{}\includegraphics[scale=0.75]{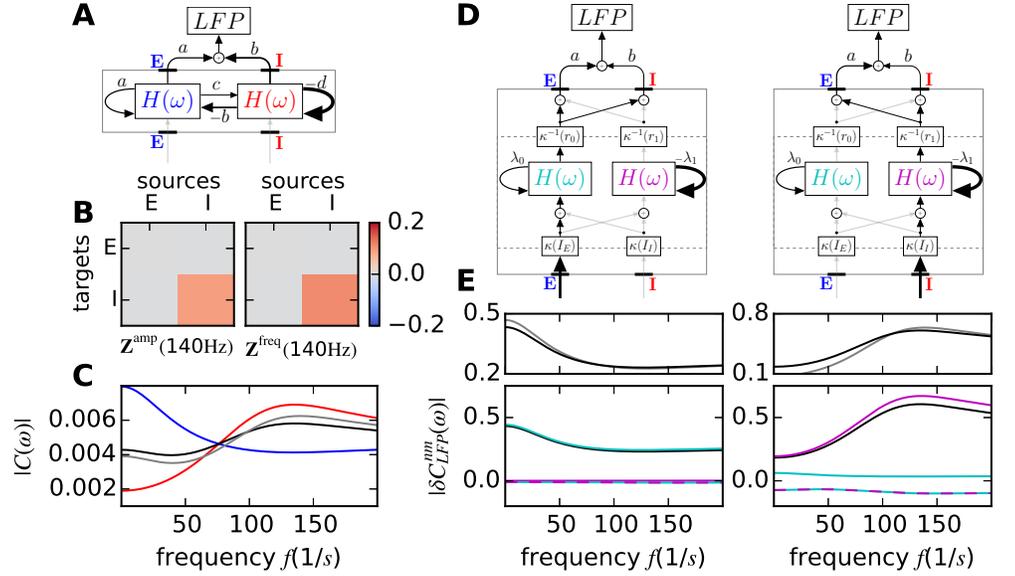}\caption{\textbf{Responses of a circuit with an embedded self-coupled inhibitory
population to oscillatory input. A} Sketch of the circuit with connectivity
defined in \prettyref{eq:conn_matrix-1}. $H(\omega)$ denotes the
dynamic transfer function of the populations (\prettyref{eq:tf_2d}).
\textbf{B }Sensitivity measure (\prettyref{eq:def_sensitivity_measure}),
showing the anatomical connections shaping the amplitude (left) and
frequency (right) of the peak in the spectrum. \textbf{C} Spectra
produced by the circuit without input. Red: spectrum of the excitatory
population $C_{\protect\E\protect\E}(\omega)$, blue: spectrum of
the inhibitory population $C_{\protect\I\protect\I}(\omega)$, gray:
LFP $C_{\mathrm{LFP}}(\omega)$ as defined in \prettyref{eq:def_LFP},
black: approximation of the LFP by the autocorrelation of the population
spectra $C_{\mathrm{LFP}}^{\mathrm{auto}}(\omega)$ as defined in
\prettyref{eq:def_LFP}. The area under is spectrum between $0\protect\Hz$
and $200\protect\Hz$ is normalized to one. \textbf{D} Sketches of
the eigenmode decomposition of the circuit and the input. The input
vector $\mathbf{I}=I_{\protect\E}\mathbf{e}_{0}+I_{\protect\I}\mathbf{e}_{1}$
is split into components applied to the individual populations $I_{\protect\E,\protect\I}$
(see arrows entering the bottom of the box). The input subsequently
undergoes a linear basis transformation $\kappa(I_{\protect\E,\protect\I})$
into the basis spanned by the eigenvectors of the effective connectivity
matrix, yielding components of the input that excite the individual
modes $\mathbf{I}=I_{0}\mathbf{u}_{0}+I_{1}\mathbf{u}_{1}$. The modes
are characterized by their transfer functions (\prettyref{eq:tf_2d})
(cyan: mode zero, magenta: mode one), which, in this case of equal
working points of the populations, are equal to the transfer function
of the populations. If the populations were set at different working
points, the transfer function of the modes would be a combined function
of the transfer functions of the populations. The rates emitted by
the modes $r_{0,1}$ are transformed back into the basis of the populations
$\kappa^{-1}(r_{0,1})$ with $\mathbf{r}=r_{0}\mathbf{u}_{0}+r_{1}\mathbf{u}_{1}=r_{\protect\E}\mathbf{e}_{0}+r_{\protect\I}\mathbf{e}_{1}$.
The arrows leaving the box at the top denote the rates of the populations
$r_{\protect\E,\protect\I}$, which are combined to the LFP signal.
The oscillatory input is applied to either the excitatory (left, $I_{\protect\I}=0$)
or the inhibitory (right, $I_{\protect\E}=0$) population. Gray arrows
denote connections with negligible strength. \textbf{E} Upper panel:
additional LFP induced by the stimulus. Full signal $\delta C_{\mathrm{LFP}}(\omega)$
(gray curve) and the signal without the crosscorrelation of the input
currents $\delta C_{\mathrm{LFP}}^{\mathrm{auto}}(\omega)$ (\prettyref{eq:def_dC_LFP_auto},
black curve). Lower panel: decomposition of the excess LFP spectra
$\delta C_{\mathrm{LFP}}^{\mathrm{auto}}(\omega)$ into the contribution
of the zeroth mode (cyan), the first mode (magenta) and the cross-modes
(cyan-magenta). The sum of all contributions (black) approximately
equals the contribution of the zeroth mode when stimulating E (left)
and the contribution of the first mode when stimulation I (right).
Connectivity parameter: $g=1.4,\:a=0.5,\ b=0.5g,\ c=0.1,\ d=g$.\label{fig:2d_ING1}}
\end{figure}

The first circuit is composed of two self-coupled populations, one
excitatory and one inhibitory. The latter is coupled to the excitatory
population, while the reverse connection is of negligible strength.
Because the network has an approximate feedforward rather than a recurrent
structure, the dynamic modes of the circuit correspond approximately
to the populations in isolation. The E-E loop generates a low pass
filter and the I-I loop a peak at around $140\Hz$ (\prettyref{fig:2d_ING1}B),
which is visible in the LFP at around $125\Hz$ (\prettyref{fig:2d_ING1}C).
This is reflected in the eigenvectors (\prettyref{eq:evecs_iso}),
which point in the direction of the populations, and in the eigenvalues
$\lambda_{0}\approx a,\ \lambda_{1}\approx-d$, which reflect the
strength of the feedback connections of the populations. Without additional
input, the signal produced by the excitatory population is dominated
by the zeroth mode. Since the self-coupling of the mode is positive,
but the dynamics are still stable, excitations decay slowly, reflected
by enhanced slow frequencies in the spectrum (\prettyref{fig:2d_ING1}C,
blue curve). The spectrum observed in the inhibitory population is
dominated by the first mode, which has a negative eigenvalue and therefore
produces an oscillation similar to the isolated populations discussed
in the previous section (\prettyref{fig:2d_ING1}C, red curve). Combining
these signals yields the LFP (\prettyref{fig:2d_ING1}C, gray and
black curve), which contains contributions of both modes. Since the
negative feedback to the inhibitory populations is stronger than the
excitatory connection, the LFP is similar to the spectrum of the inhibitory
neurons. At low frequencies, the spectrum of the excitatory population
is particularly large and therefore raises the LFP signal. 

Stimulating the excitatory population excites only the zeroth mode
($\gamma_{0}=\mathbf{v}_{0}^{\mathrm{T}}\mathbf{e}_{0}\approx1,\ \gamma_{1}\approx0$),
as sketched in the left panel of \prettyref{fig:2d_ING1}D and reflected
in the additional LFP response (\prettyref{fig:2d_ING1}E, left panels).
Similarly, stimulating the inhibitory population excites the first
mode (\prettyref{fig:2d_ING1}D, right) and therefore yields a high
frequency peak in the LFP response (\prettyref{fig:2d_ING1}E, right). 

In summary, in a two-dimensional network, with a feedforward structure
from the inhibitory to the excitatory population, each population
generates its own rhythm by self-coupling. As a result, the excitation
of individual populations elicits responses which can be traced back
to the original circuits generating the oscillations observable in
the LFP. In particular, stimulating the inhibitory population yields
a high frequency peak in the additional LFP spectrum, stimulation
the excitatory population, on the other hand, yields increased low
frequencies. 

\begin{figure}[h]
\centering{}\includegraphics[scale=0.8]{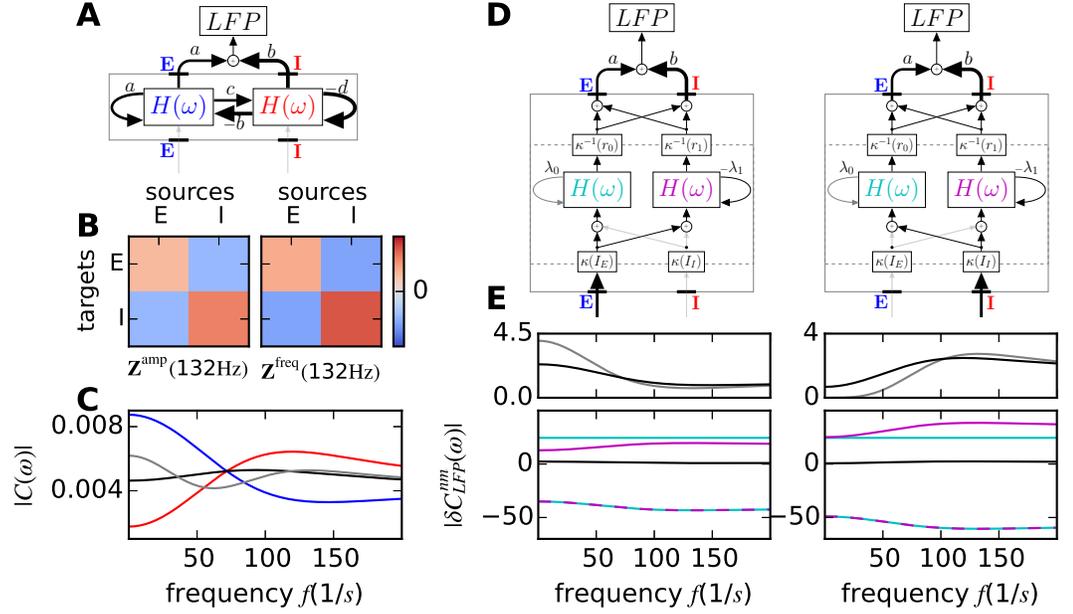}\caption{\textbf{Responses of a symmetric two-dimensional circuit. A} Sketch
of the circuit. \textbf{B }Sensitivity measure (\prettyref{eq:def_sensitivity_measure}),
showing the anatomical connections shaping the amplitude and frequency
of the peak in the spectrum. \textbf{C} Spectra produced by the circuit
without input. Red: spectrum of the excitatory population $C_{\protect\E\protect\E}(\omega)$,
blue: spectrum of the inhibitory population $C_{\protect\I\protect\I}(\omega)$,
gray: LFP $C_{\mathrm{LFP}}(\omega)$ as defined in \prettyref{eq:def_LFP},
black: approximation of the LFP by the autocorrelation of the population
spectra $C_{\mathrm{LFP}}^{\mathrm{auto}}(\omega)$ as defined in
\prettyref{eq:def_LFP}. The area under is spectrum between $0\protect\Hz$
and $200\protect\Hz$ is normalized to one. \textbf{D} Sketches of
the eigenmode decomposition of the circuit and the input to the populations.
The eigenmodes are characterized by their transfer functions (cyan:
mode zero, magenta: mode one). The oscillatory input is applied to
either the excitatory (left) or inhibitory (right) population. Gray
arrows denote connections with negligible strength. \textbf{E} Upper
panel: additional LFP induced by the stimulus. Full signal $\delta C_{\mathrm{LFP}}(\omega)$
(gray curve) and the signal without the crosscorrelation of the input
currents $\delta C_{\mathrm{LFP}}^{\mathrm{auto}}(\omega)$ (\prettyref{eq:def_dC_LFP_auto},
black curve). Lower panel: decomposition of the excess LFP spectra
$\delta C_{\mathrm{LFP}}^{\mathrm{auto}}(\omega)$ into the contribution
of the zeroth mode (cyan), the first mode (magenta) and the cross-modes
(cyan-magenta). Connectivity parameter: $g=1.4,\:a=1.0,\ b=g,\ c=1.0,\ d=g$.\label{fig:2d_ING2}\textbf{}}
\end{figure}

\subsubsection*{Symmetric two-dimensional network\label{subsec:ING2}}

A peak in the high $\gamma$-range has been shown to be generated
by a network composed of LIF-neuron models with symmetric architecture
\citep{Brunel00}. In this setup the excitatory and the inhibitory
population receive the same input ($a=c=1,\ b=d=g$) (\prettyref{fig:2d_ING2}A)
yielding one eigenvalue to be zero and one to be negative ($\lambda_{0}=0,\ \lambda_{1}=1-g$),
for networks working in the inhibition dominated regime ($g>1$).
Here the eigenvectors deviate from the vectors of the populations
($\mathbf{e}_{0}$ and $\mathbf{e}_{1}$), revealing that the modes
get contributions from both populations. Since the zeroth eigenvalue
is zero, the corresponding mode has no feedback (\prettyref{fig:2d_ING2}D,
left) and the produced spectrum, which scales with $1/\left|1-\lambda_{0}H(\omega)\right|$
(\prettyref{eq:spec_2d}), is therefore constant. The first mode generates
a peak due to the negative feedback, with a frequency of around $130\Hz$
determined by the parameters of its transfer function as well as the
strength of the coupling $\lambda_{1}$. The sensitivity analysis
shows that this peak is generated by an interplay of the two populations
as well as their self-coupling (\prettyref{fig:2d_ING2}B). 

The LFP produced by the circuit without additional input, as well
as its decomposition into the spectra observed in the populations,
is shown in \prettyref{fig:2d_ING2}C. Since the population vector
of the inhibitory population points more in the direction of the right
eigenvector of the first mode than the right eigenvector of the zeroth
mode ($\mathbf{u}_{1}^{\mathrm{T}}\mathbf{e}_{1}>\mathbf{u}_{0}^{\mathrm{T}}\mathbf{e}_{1}$),
the spectrum of the inhibitory population displays the peak generated
by the first mode. The opposite is true for the excitatory population
($\mathbf{u}_{0}^{\mathrm{T}}\mathbf{e}_{0}>\mathbf{u}_{1}^{\mathrm{T}}\mathbf{e}_{0}$)
which shows a spectrum dominated by the zeroth mode and the mixture
of the zeroth and first mode, which is reminiscent of a low pass filter.
Note, however, that this reasoning only holds for modes which are
far from an instability. If one of the eigenvalues would approach
the value one at a certain frequency, the prefactor in \prettyref{eq:spec_2d}
would become large and the corresponding peak would be visible in
all populations. However, in the presented regime of weakly synchronized
oscillations, the rhythm generated by the full circuit produces similar
population rate spectra as the rhythm generated by a single population
(compare \prettyref{fig:2d_ING2}C and \prettyref{fig:2d_ING1}C).
It is notable, that in this circuit the crosscorrelation of the population
rate spectra have a large impact on the LFP (\prettyref{fig:2d_ING2}C
).

Applying a stimulus to one of the two populations elicits large responses
originating in the autocorrelations of the modes, as well as in their
crosscorrelation (see single color and dashed curves in \prettyref{fig:2d_ING2}E).
The large crosscorrelation can be understood by considering that the
same connections contribute to both modes. That was not the case in
the previous circuit, where the crosscorrelation therefore remained
small. In the current circuit, the sum of the auto- and crosscorrelation
is small and thus the resulting spectrum is also small. An example
of such a circuit is a network of neurons where the population signal
displays small fluctuations, but a projection of the rates onto the
direction of the modes yields strongly fluctuating signals, which
are highly negatively correlated between the modes.

In quantitative terms, stimulating the excitatory population elicits
responses of the modes whose amplitudes scale with $\gamma_{0}=\frac{\sqrt{g^{2}+1}}{g-1}$
and $\gamma_{1}=-\frac{\sqrt{2}}{g-1}$ (\prettyref{fig:2d_ING2}D,
left). The signs of $\gamma_{0}$ and $\gamma_{1}$ reveal that
excitation of the two modes is of opposite signs, yielding a negative
crosscorrelation (which scales with $\gamma_{1}\gamma_{2}$, see \prettyref{fig:2d_ING2}E,
left). As the population vector of the excitatory population points
more in the direction of the right eigenvector corresponding to the
zeroth mode than the eigenvector of the first mode ($|\gamma_{0}|>|\gamma_{1}|$),
the stimulus induced response visible in the LFP is dominated by contributions
of the zeroth mode and the coupling of the zeroth and first mode (\prettyref{fig:2d_ING2}E
left), as already observed in the composition of the spectrum without
additional stimulus. Stimulating the inhibitory population also excites
both modes ($\gamma_{0}=-\frac{\sqrt{g^{2}+1}}{g-1},\ \gamma_{1}=\frac{g\sqrt{2}}{g-1}$)
with opposite signs (\prettyref{fig:2d_ING2}D, right) such that the
main part of the responses cancel. The contribution of the first mode
is slightly larger (\prettyref{fig:2d_ING2}E, right), imposing its
peak onto the LFP spectrum. 

Thus the responses to stimulations obtained here are similar to the
responses of the previously considered circuitry: stimulating the
excitatory population elicits mainly slow frequencies, while stimulating
the inhibitory population reveals a high frequency peak. These responses
can therefore not distinguish between an I-I-loop and a fully connected
E-I-circuit generating the high frequency peak.

In principle, it is possible to selectively probe the modes of a complex
circuit experimentally. This requires co-stimulation of all populations
with population specific stimulus amplitudes chosen proportional to
their respective entries in the right-sided eigenvectors $\mathbf{u}_{0}$
and $\mathbf{u}_{1}$.

\subsubsection*{A two-dimensional network without self-coupling\label{subsec:An-two-dimensional-network}}

Recurrently connected excitatory and inhibitory populations can produce
oscillations without the necessity of inhibitory self-feedback. The
network motif discussed in this section can be considered the prototypical
PING motif as discussed in \citep{Kang09_1573}, which produces $\gamma$
oscillations.

\begin{figure}[htp]
\centering{}\includegraphics[scale=0.8]{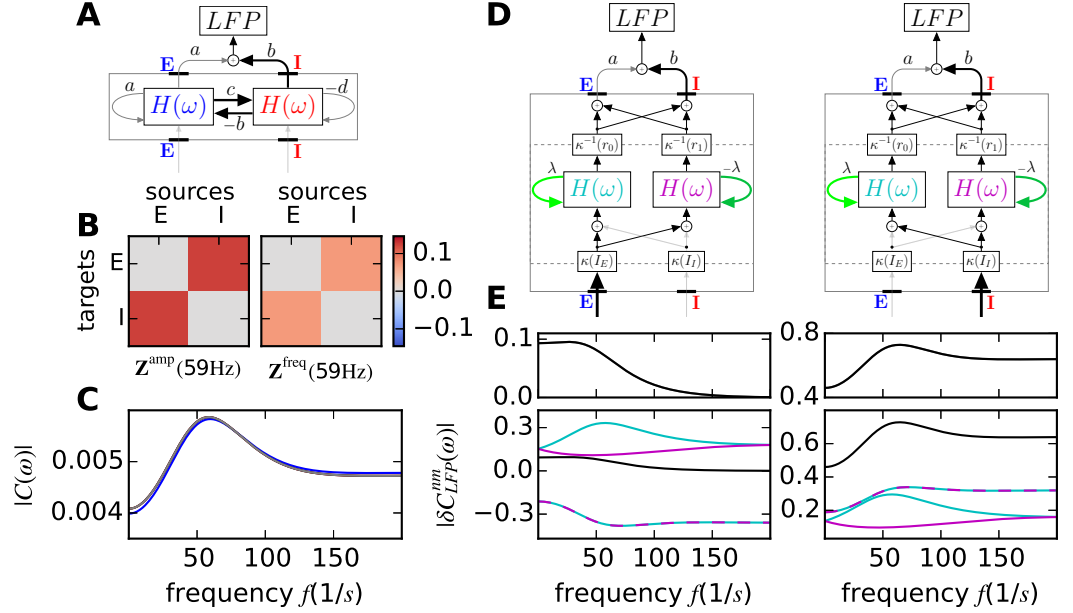}\caption{\textbf{Response to oscillatory input of a two-dimensional circuit
without self-couplin}g\textbf{. A} Sketch of the circuit. \textbf{B
}Sensitivity measure (\prettyref{eq:def_sensitivity_measure}), showing
the anatomical connections shaping the amplitude and frequency of
the peak in the spectrum. \textbf{C} Spectra produced by the circuit
without input. Red: spectrum of the excitatory population $C_{\protect\E\protect\E}(\omega)$,
blue: spectrum of the inhibitory population $C_{\protect\I\protect\I}(\omega)$,
gray: LFP $C_{\mathrm{LFP}}(\omega)$ as defined in \prettyref{eq:def_LFP},
black: approximation of the LFP by the autocorrelation of the population
spectra $C_{\mathrm{LFP}}^{\mathrm{auto}}(\omega)$ as defined in
\prettyref{eq:def_LFP}. The area under is spectrum between $0\protect\Hz$
and $200\protect\Hz$ is normalized to one. \textbf{D} Sketches of
the eigenmode decomposition of the circuit and the input to the populations.
The eigenmodes are characterized by their transfer functions (cyan:
mode zero, magenta: mode one). Green denotes positive complex values
($\lambda_{0}=i\sqrt{bc}$) and dark green negative complex values
($\lambda_{1}=-i\sqrt{bc}$). The oscillatory input is applied to
either the excitatory (left) or inhibitory (right) population. Gray
arrows denote connections with negligible or zero strength. \textbf{E}
Upper panel: additional LFP induced by the stimulus. Full signal $\delta C_{\mathrm{LFP}}(\omega)$
(gray curve) and the signal without the crosscorrelation of the input
currents $\delta C_{\mathrm{LFP}}^{\mathrm{auto}}(\omega)$ (\prettyref{eq:def_dC_LFP_auto},
black curve). Lower panel: decomposition of the excess LFP spectra
$\delta C_{\mathrm{LFP}}^{\mathrm{auto}}(\omega)$ into the contribution
of the zeroth mode (cyan), the first mode (magenta) and the cross-modes
(cyan-magenta). Connectivity parameter: $a=0,\ b=0.8,\ c=0.9,\ d=0$.\label{fig:2d_PING}}
\end{figure}

\prettyref{fig:2d_PING}A shows a diagram of a circuit with connections
between the populations and negligible self-couplings. In this parameter
regime the LFP is determined by the inhibitory activity impinging
onto the excitatory neurons. This circuit generates a peak at around
$50\Hz$, which, as expected, depends on the connections between the
populations (\prettyref{fig:2d_PING}B). Since the eigenmode producing
the oscillations is a mixture of the two populations, with both of
them having comparably sized entries in the corresponding eigenvectors,
the population spectra are similar in both populations with similar
contribution to the LFP (\prettyref{fig:2d_PING}C, all curves lie
on top of each other). The circuit is characterized by two eigenvalues,
which are complex conjugates and purely imaginary ($\lambda_{0,1}=\pm i\sqrt{bc}$).
Considering the eigenvalue trajectories (described by $\lambda_{0,1}H(\omega)$)
reveals that the trajectory starting at a positive imaginary part
produces the $50\Hz$ peak, while the other trajectory produces a
peak at a very large frequency. The latter one is potentially suppressed
in neural circuits by inhomogeneities in the parameters, like distributed
delays, which contribute an additional multiplicative factor with
low-pass characteristics (cf. \prettyref{eq:def_H}) to the transfer
function. We hence focus on the peak at lower frequency.

A stimulus applied to either the excitatory or the inhibitory population
excites both modes with equal strength ($\gamma_{0}=\gamma_{1}=\frac{1}{2}\sqrt{\frac{bc+1}{bc}}$
for stimulation of the excitatory population and $\gamma_{0}=\gamma_{1}^{*}=\frac{i}{2}\sqrt{bc+1}$
for stimulation of the inhibitory population,  illustrated in \prettyref{fig:2d_PING}D).
Since $\gamma_{0}\gamma_{1}^{*}>0$ when stimulating the excitatory
and $\gamma_{0}\gamma_{1}^{*}<0$ when stimulating the inhibitory
population, the contribution of the crosscorrelations to the response
spectrum is of opposite sign for the two stimuli, as seen in \prettyref{fig:2d_PING}E
(dashed curves). The cancellations of the cross- and autocorrelation
when stimulating the excitatory population yields an LFP response,
which is reminiscent of a low-pass filter with a small peak at very
low frequencies (\prettyref{fig:2d_PING}E, left upper panel). Since
no cancellation occurs when stimulating the inhibitory population,
the spectrum shows amplifications at the frequency that is generated
by the circuit autonomously (\prettyref{fig:2d_PING}E, right upper
panel). Hence, also with this circuit motif, stimulation of the inhibitory
population yields a peak in the LFP which is missing when stimulating
the excitatory population, even though the excitatory population is
involved in generating the peak.

In order to isolate the response of one mode, the stimulus vector
needs to point in the direction of its right eigenvector. Since the
eigenvectors are complex, with real entries for the excitatory and
imaginary entries for the inhibitory population ($\mathbf{u}_{0,1}=\sqrt{\frac{1}{bc+1}}\left(\sqrt{bc},e^{\mp i\pi/2}\right)$),
adjusting the amplitude of the sinusoidal signal applied to the population
is not sufficient to segregate the mode responses. Since the Fourier
transform of a phase-shifted sine wave is given by $\mathcal{F}[sin(\omega_{0}t+\phi)]=e^{i\omega/\omega_{0}\phi}\mathcal{F}[\sin(\omega_{0}t)]$,
complex entries in the stimulus vector can be achieved by adjusting
the relative phase of the input to the populations. The mode generating
the peak around $50\Hz$ is excited in isolation if the stimulus applied
to the inhibitory population lags the stimulus to the excitatory population
by $\pi/2$. Reversely, if the excitation of the inhibitory population
precedes the stimulation of the excitatory population by $\pi/2$,
the LFP response is determined by the first mode, which amplifies
fast oscillations.

\subsection*{Stimulus evoked spectra in a model of a microcircuit}

In this section we analyze the responses observed in a model of a
microcircuit of the primary sensory cortex to oscillatory input, discuss
the results in comparison with experimental data and point out potential
pitfalls when utilizing these results to identify the anatomical sources
of oscillations produced by the circuit. First, a previously introduced
theoretical framework, which enables the prediction of population
rate spectra as well as the location of their origin, is extended
by oscillatory input. Subsequently we demonstrate that the theoretical
prediction reproduces the responses observed in simulations and additionally
offers insight into the anatomical origin of the components contributing
to these responses. Analyzing the responses of the populations in
layer 2/3, we demonstrate that ad-hoc interpretations can yield misleading
conclusions.

\subsubsection*{The microcircuit with oscillatory input}

The microcircuit model has been introduced by Potjans et al. \citep{Potjans14_785}
to represent a layered circuit typical for the primary sensory cortex.
The model is composed of around $10^{5}$ LIF model neurons, which
are divided into four layers (L2/3, L4, L5, L6) with one excitatory
and one inhibitory population each. Connection probabilities between
theses eight populations are gathered from 50 anatomical and physiological
studies. The model has been shown to reproduce typical rate profiles
\citep{Potjans14_785}. In agreement with experimental evidence it
supports the emergence of slow rate fluctuations in layer 5, as well
as low- and high-$\gamma$ oscillations in the upper layers \citep{Bos16_1}.
As yet, the microcircuit has only been analyzed in the resting state,
when each population receives uncorrelated Poisson input, which mimics
the input of remote areas. Oscillatory stimuli are introduced by modulating
a ratio $a$ ($0<a<1$) of the external input rate to the populations
with a sinusoid of a given frequency $f_{\mathrm{ext}}$

\begin{equation}
\nu_{\mathrm{ext}}^{k}(t)=\nu_{\mathrm{ext}}^{0}N_{\mathrm{ext}}^{k}(1+a\,\sin(2\pi f_{\mathrm{\mathrm{ext}}}t)).\label{eq:modulated-ext-rate}
\end{equation}
Here $\nu_{\mathrm{ext}}^{k}(t)$ denotes the total external input
applied to the $k$-th population, $\nu_{\mathrm{ext}}^{0}$ the firing
rate associated with one incoming connection and $N_{\mathrm{ext}}^{k}$
the external indegree to population $k$. 

\begin{figure}[ht]
\begin{centering}
\includegraphics[scale=0.75]{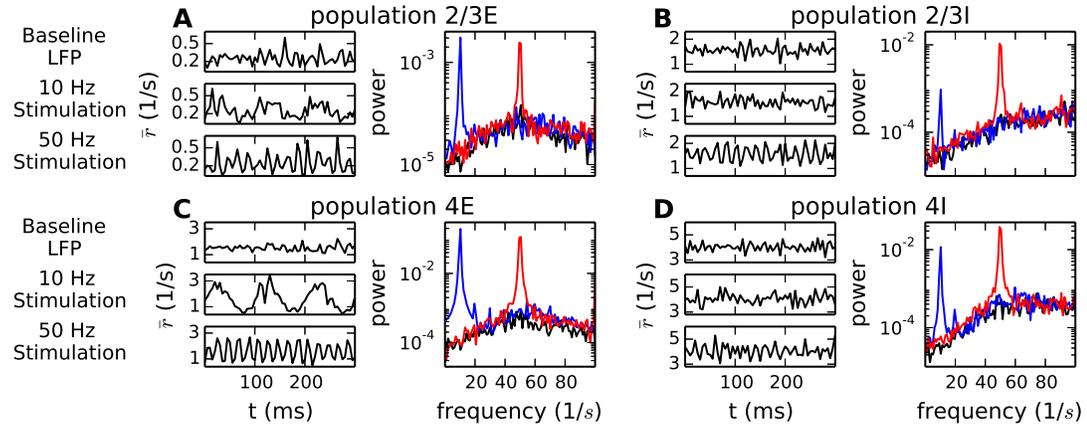}\caption{\textbf{Population rate responses to oscillatory stimuli. A }Left:
Instantaneous firing rate of population 2/3E of the microcircuit in
the resting state (top) and with additional sinusoidal stimulus (\prettyref{eq:modulated-ext-rate})
of frequency $10\protect\Hz$ (middle) and $50\protect\Hz$ (bottom)
applied to population 2/3E. Right: Corresponding power spectra of
the rates in population 2/3E in the resting state (black) and with
additional stimuli ($10\protect\Hz$: blue, $50\protect\Hz$: red).
\textbf{B}-\textbf{D }Instantaneous firing rates and power spectra
of populations 2/3I, 4E, and 4I as in A, with the stimuli applied
to the respective population.\label{fig:Population-rate-responses} }
\par\end{centering}
\end{figure}

\prettyref{fig:Population-rate-responses} shows the instantaneous firing rate, as well as the spectra observed
in populations 2/3E, 2/3I, 4E, and 4I in the resting state and with
oscillatory input of $10\Hz$ and $50\Hz$ to the populations of strength
$a=0.05$. The spectra show that this modulation of $5$ percent of
the external static input  suffices to reproduce population responses
of strength comparable to LFP responses measured in experiments applying
oscillatory light stimuli to optogenetically altered mice (see Fig.
3c in \citep{Cardin2009}). The rate of population 4E adapts strongly
to both rhythms (see left panel of \prettyref{fig:Population-rate-responses}C),
which is interesting since layer 4 is considered the main recipient
of thalamic input \citep{Potjans14_785}. In the microcircuit, excitatory
populations show a stronger amplification of the $10\Hz$ stimulus,
while the inhibitory populations show higher amplitudes when stimulated
with $50\Hz$ (\prettyref{fig:Population-rate-responses}). These
results agree with measurements conducted by Cardin et al. (Supplementary
Fig. 5 in \citep{Cardin2009}) by trend: Low frequency stimulation
of excitatory neurons show stronger responses in the LFP than stimulation
of inhibitory neurons and vice versa for high frequency stimulations.
However, both excitatory and inhibitory currents contribute to the
LFP \citep{Mazzoni2015}, such that the measured responses in experiments
should be compared with a superposition of the individual population
rate spectra. 

The effect that low frequency stimulation of excitatory populations
evokes strong responses can be understood intuitively: Our previous
examples show that excitatory populations more prominently participate
in modes that resemble low-pass filters. In other words, excitatory
activity and, in particular, excitatory self-feedback drive the circuit
towards a rate instability which facilitates slowly decaying modes.
Responses of these modes are also elicited when stimulating the excitatory
population at low frequencies, resulting in amplified low frequency
responses.

\subsubsection*{Theoretical description of oscillatory input}

Here we describe how the amplification of oscillatory stimuli can
be understood theoretically. Previous work \citep{Bos16_1} shows
that the population rate spectra produced by this model are sufficiently
described by a theoretical two-step reduction composed of mean-field
theory, which yields the stationary firing rates \citep{Amit97,Fourcaud02},
as well as linear response theory, which characterizes the response
properties of the neurons \citep{Brunel00,Brunel99,Schuecker15_transferfunction},
formally described by the dynamic transfer function. Since already
small modulations of the external input have considerable impact on
the population rate spectra, but negligible effect on the stationary
firing rates (\prettyref{fig:Population-rate-responses}), we constrain
our analysis in this section to purely oscillatory stimuli which do
not alter the working point of the populations. This assumption can
be justified by the observation that the width of the response peaks
shown in \prettyref{fig:Population-rate-responses} is narrow (in
particular for high frequencies) and the stimulated frequency therefore
approximately does not couple to other frequencies. The network can
therefore be analyzed analogously to the two-dimensional circuits
discussed in the previous section, after mapping the dynamics in the
microcircuit model to interacting linear rate models \citep{Bos16_1}.
Extensions to stimuli that affect the stationary state of the system
are discussed in \nameref{subsec:Approximation_of_dtf}. The spectrum
as well as the excess spectrum $\delta\mathbf{C}(\omega)$ of the
microcircuit with sinusoidal input are hence described by the same
equation as the 2d-circuit (\prettyref{eq:spec_2d}, \prettyref{eq:excess_spec_2d})
extended to eight populations (for detail see \nameref{subsec:spec_input})

\begin{align}
\delta\mathbf{C}(\omega)=\mathbf{C}_{I}(\omega)-\mathbf{C}(\omega) & =\sum_{n,m=1}^{8}\frac{\beta_{\mathrm{ext},nm}(\omega)}{(1-\lambda_{n}(\omega))(1-\lambda_{m}^{*}(\omega))}\mathbf{u}_{n}(\omega)\mathbf{u}_{m}^{\mathrm{T}*}(\omega),\label{eq:spectrum_decomp_with_input_microcircuit}
\end{align}
with (\prettyref{eq:beta_ext})
\begin{equation}
\beta_{\mathrm{ext},nm}(\omega)=w_{\mathrm{ext}}^{2}N_{\mathrm{ext},k}^{2}\nu_{\mathrm{ext}}^{2}a^{2}\frac{T}{4}\left|H_{k}(\omega_{\mathrm{I}})\right|^{2}\alpha_{n}^{k}(\omega)\alpha_{m}^{k*}(\omega).\label{eq:beta_ext_text}
\end{equation}
The latter factor, which quantifies the amplitude of the excited auto
and cross-correlations of the modes, depends on the modulated external
firing rates $N_{\mathrm{ext},k}\nu_{\mathrm{ext}}a$, the external
synaptic weights $w_{\mathrm{ext}}$, the transfer function of the
population that receives the input $H_{k}(\omega_{\I})$, the projection
of the modes on the direction of the stimulus $\alpha_{n}^{k}(\omega)=\mathbf{v}_{n}^{\mathrm{T}}(\omega)\mathbf{e}_{k}$
as well as the measurement time $T$.

\begin{figure}[ht]
\centering{}\includegraphics[scale=0.75]{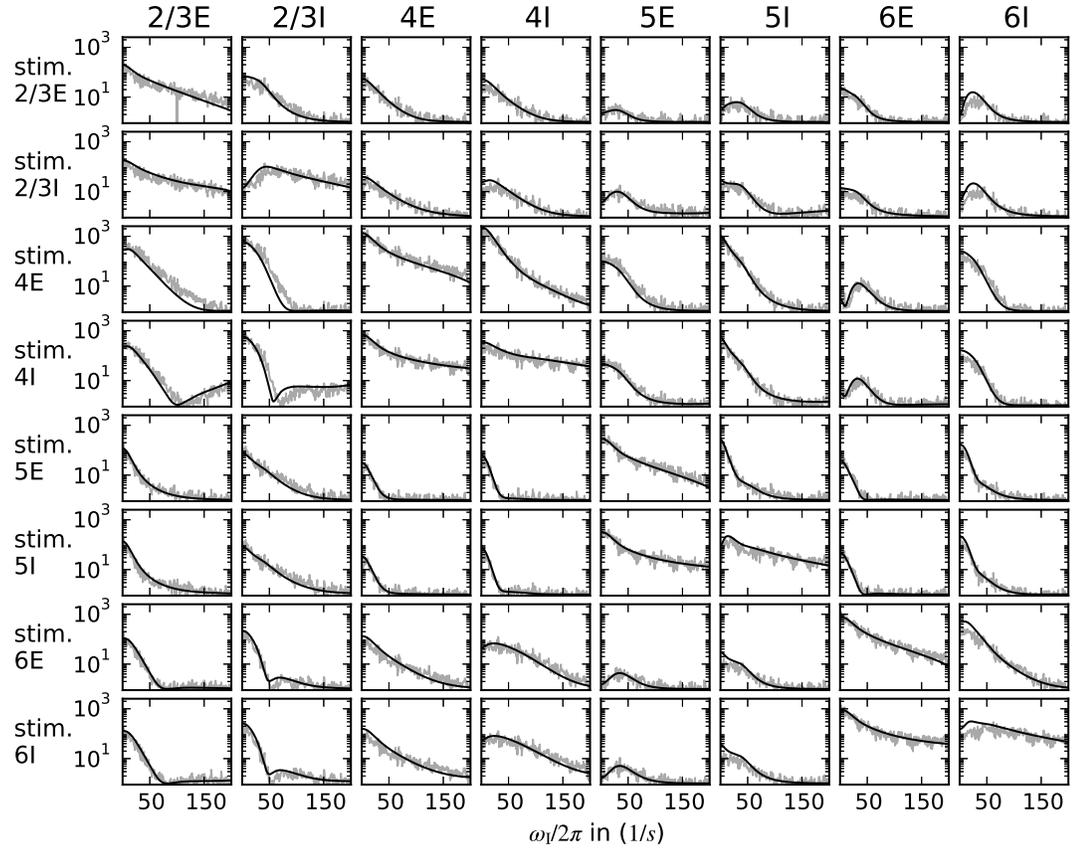}\caption{\textbf{Power ratios of population rate spectra with oscillatory stimuli.
}Each column shows the power ratios (\prettyref{eq:power_ratio_microcircuit})
of one population (gray: simulation, black: theoretical prediction).
The rows indicate the population the stimulus is applied to. The stimulus
is applied as a sinusoidal modulation of $5$ percent of the external
static input rate and oscillates with frequencies between $0$ and
$200\protect\Hz$ in steps of $1\protect\Hz$. The power ratio is
measured at the frequency of the stimulus.\label{fig:Power-ratios-sim}}
\end{figure}

Since the populations are set at different working points, which is
reflected in different  population specific firing rates and transfer
functions, the left and right eigenvectors are frequency dependent,
in contrast to the models considered in the previous sections. The
power ratio, which describes the relative size of the evoked spectrum
at stimulation frequency, is given by

\begin{equation}
\rho(\omega)=\frac{\mathbf{C}_{I}(\omega)}{\mathbf{C}(\omega)}=1+\frac{\delta\mathbf{C}(\omega)}{\mathbf{C}(\omega)}\label{eq:power_ratio_microcircuit}
\end{equation}
and evaluated at stimulus frequency $\omega=\omega_{I}$. To show
that the theoretical description suffices to predict the impact of
oscillatory input to the population rate spectra, we modulate $5\,$percent
of the static external input to each population at frequencies between
$0$ and $200\Hz$ and compare the power ratios at stimulation frequency
observed in each population with the theoretically predicted power
ratios (\prettyref{fig:Power-ratios-sim}). As expected, the response
to a stimulus is strongest in the population the stimulus is applied
to (see the diagonal panels in \prettyref{fig:Power-ratios-sim}).

Considering only the diagonal panels, it is evident that stimuli applied
to excitatory populations amplify mostly low frequencies, which confirms
the previously found tendencies. The inhibitory populations, except
for 4I, display resonances at frequencies larger than zero. The most
prominent peak is visible in population 2/3I, which, due to its location
at just above $40\Hz$, could be interpreted as a resonance phenomenon
related to the low-$\gamma$ peak produced by the circuit. The responses
of 2/3E and 2/3I (\prettyref{fig:Composition-of-power}A0, B0) show
similar tendencies as the LFP power ratio measured in experiments
(see Fig. 3 in \citep{Cardin2009}), namely 2/3E supports mainly low
frequencies, while 2/3I displays a resonance at around $41\Hz$. However,
it remains to be investigated whether the peaks in the population
rate spectra would also be visible in the LFP, which is given as a
weighted sum of the spectra of one row in \prettyref{fig:Power-ratios-sim}
in addition to the crosscorrelations of the currents (as outlined
for an examplary 2d-circuit in \nameref{subsec:LFP}). 

\subsubsection*{The origin of the peak visible in the population rate spectrum of
2/3I}

\begin{figure}[h]
\centering{}\includegraphics[scale=0.75]{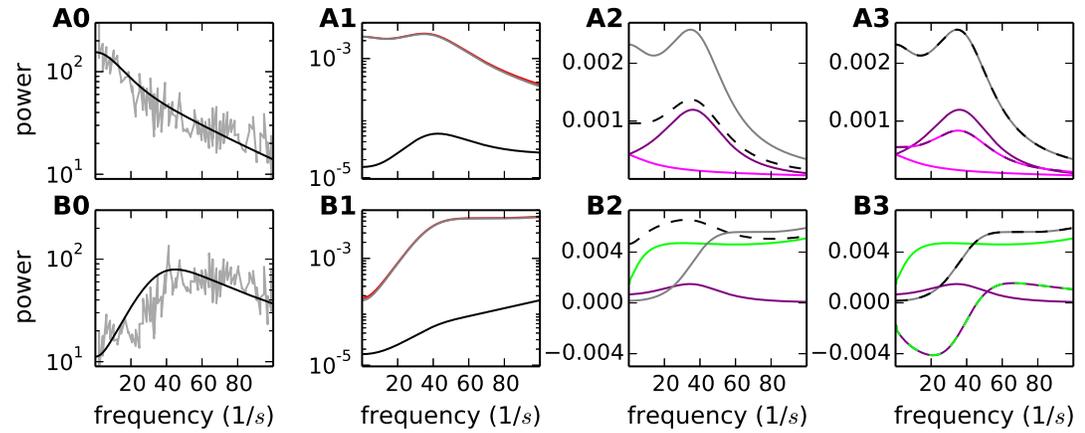}\caption{\textbf{Composition of power ratio in populations 2/3E and 2/3I. A0
}Power ratio of population 2/3E with oscillatory input to 2/3E (gray:
simulation, black: theoretical prediction, see \prettyref{fig:Power-ratios-sim}).
\textbf{A1 }Population rate spectrum of population 2/3E in the resting
state (black, $\mathbf{C}_{0}(\omega)$) and with oscillatory input
(red, $\mathbf{C}_{\protect\I}(\omega)$, \prettyref{eq:spectrum_decomp_with_input}).
The gray curve indicates the additional power induced by the stimulus
(excess spectrum: $\mathbf{C}_{\protect\I}(\omega)-\mathbf{C}_{0}(\omega)$).
\textbf{A2 }Contributions due to the autocorrelation of the modes
(see \prettyref{eq:spectrum_decomp_with_input_microcircuit} for $n=m$,
colored curves) to the excess spectrum (gray curve). Different colors
correspond to different modes, with their anatomical origin shown
in \prettyref{fig:Anatomical-origins-of}. Only the dominant contributions
are shown. Each mode can be attributed to the generation of (at most)
one peak (for the modes shaping the oscillations in the microcircuit
see Methods in \citep{Bos16_1} and \prettyref{fig:Anatomical-origins-of}).
The sum of all contributions originating in the autocorrelations of
the modes is depicted by the black dashed curve. \textbf{A3 }As in
A2 including the contribution originating in the crosscorrelation
of the modes (see \prettyref{eq:spectrum_decomp_with_input_microcircuit}
for $n\protect\neq m$). The identities of the modes contributing
to the pairwise contributions are reflected in the color-composition
of the curves. The sum of all contributions (black dashed curve) is
identical to the entire excess spectrum (gray curve).\textbf{ B0}-\textbf{B3
}Same as in A0-A3, for the population rate dynamics observed in population
2/3I with oscillatory input to population 2/3I. The areas underneath
the curves of the sum of all modes deviate from the area underneath
the curve of the sum of the two shown modes by 4\% (A2), 7\% (B2),
21\% (A3) and 20\% (B3).\label{fig:Composition-of-power}}
\end{figure}
In the following, the results of the last sections are exploited to
analyze whether and how the peak in the power ratio of population
2/3I can shed light on the circuit producing the low-$\gamma$ oscillation.
Since power ratios can give misleading results, especially if the
resting spectra are not entirely flat (as discussed in \nameref{subsec:I-I-loop}),
the resting state spectra as well as the response and excess spectra
are shown in \prettyref{fig:Composition-of-power}A1 and \prettyref{fig:Composition-of-power}B1.
The response spectra are between one to two orders of magnitude larger
than the resting state spectra and are therefore approximately equivalent
to the excess spectra. The excess spectrum for population 2/3E shows
a peak in the low-$\gamma$ frequency range, as well as a large offset
for low frequencies. Since the stimulus amplifies low frequencies
stronger than the low-$\gamma$ peak, the $\gamma$ peak is not visible
in the power ratio. The excess spectrum visible in 2/3I has a contrary
tendency: it is weak for low frequencies and saturates at a higher
value for frequencies above $40\Hz$. The peak in the power ratios
originates in the prompt increase of the excess spectrum between $20$
and $40\Hz$.

In the following we decompose the excess spectrum into contributions
from the respective dynamic modes (see \nameref{eq:spectrum_decomp_with_input_microcircuit})
and identify the dominant contributions at peak frequency $41\Hz$.
Subsequently we employ the recently derived sensitivity measure \citep{Bos16_1}
to trace the dominant modes back to their anatomical origin.

\begin{figure}[h]
\centering{}\includegraphics[scale=0.75]{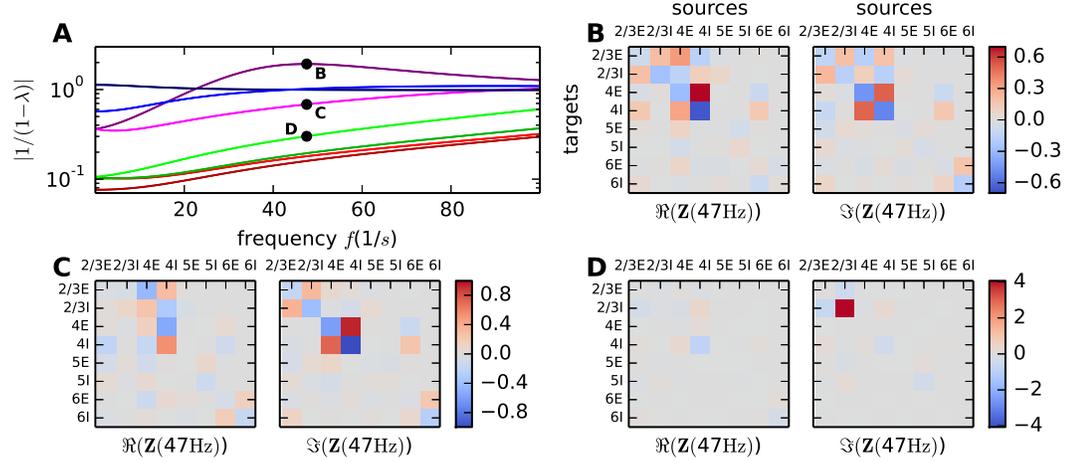}\caption{\textbf{Anatomical origins of the modes contributing to the response
spectra. A }Frequency dependence of the factors $\left|1/(1-\lambda_{i}(\omega))\right|$
that shape the population rate spectra. Each of the eight factors
is associated with a sub-set of connections, which varies with frequency.
The dominant contributions to the responses of population 2/3E and
2/3I are given by the modes whose eigenvalues are associated to the
purple, pink, and green curves. The peak frequency is marked by solid
black dots, alongside with a letter denoting the panel that shows
the associated connections. \textbf{B} Real and imaginary part of
the sensitivity measure (\prettyref{eq:def_Z}) evaluated at $47\protect\Hz$.
Strong colors reflect connections which are strongly present in the
corresponding mode. \textbf{C},\textbf{ D} Anatomical origin of pink
and green curves, as in B.\label{fig:Anatomical-origins-of}}
\end{figure}

We recall from the previous chapters, that the spectrum visible in
one population is composed of contributions arising from the autocorrelation
as well as the crosscorrelation of the modes. If the peak in the power
ratio of population 2/3I would reflect the resonance of one sub-circuit,
one would expect the excess spectrum to be composed mainly of the
contribution of the corresponding mode. The diagonal contributions
to the excess spectrum are given by

\begin{equation}
\delta\mathbf{C}_{\mathrm{auto}}(\omega)=\sum_{n=1}^{8}\frac{\beta_{\mathrm{ext},nn}(\omega)}{\left|1-\lambda_{n}(\omega)\right|^{2}}\,\mathbf{u}_{n}(\omega)\mathbf{u}_{n}^{T*}(\omega).\label{eq:spec_diag}
\end{equation}
The dominant contributions originating in the autocorrelations of
the modes as well as the sum of all contributions, are shown in \prettyref{fig:Composition-of-power}A2
and \prettyref{fig:Composition-of-power}B2. The main contribution
to the low-$\gamma$ peak in the excess spectrum of population 2/3E
is indeed given by the trajectory corresponding to the origin of the
low-$\gamma$ peak (compare dashed to purple curve), which is determined
by connections located in layer 2/3 and 4 (\prettyref{fig:Anatomical-origins-of}B).
The eigenvalue trajectory, that gives rise to the low-$\gamma$ peak,
originates in a pair of complex conjugate eigenvalues (see \prettyref{fig:Anatomical-origins-of}A
at frequency zero, where their absolute values agree), similar to
the eigenvalues in the exemplary circuit where the rhythm was entirely
determined by the coupling between the excitatory and the inhibitory
population rather than their self-coupling (see \nameref{subsec:An-two-dimensional-network}).
Hence the modes corresponding to the complex conjugated pair of eigenvalues
are shaped by the same connections and a stimulus applied to population
2/3E also elicits a response of the mode associated to the counterpart
of the low-$\gamma$ eigenvalue (see pink trajectory, its origin \prettyref{fig:Anatomical-origins-of}C
and its contribution to the excess spectrum \prettyref{fig:Composition-of-power}A2).
However, even though the peak shape appears to be formed by these
modes, the offset of the excess spectrum cannot be explained by the
diagonal contributions alone (compare dashed to gray curve). Population
2/3I contributes to the generation of two different peaks, the low-$\gamma$
oscillation, as well as a high-frequency oscillation originating in
the self-coupled population alone (see green eigenvalue trajectory
in \prettyref{fig:Anatomical-origins-of}A and its origin in \prettyref{fig:Anatomical-origins-of}D).
Hence, stimulating population 2/3I elicits responses of modes with
different anatomical origins (\prettyref{fig:Composition-of-power}B2).
The sum of contributions originating in the autocorrelations, however,
deviates considerably from the total excess spectrum. This shows that
crosscorrelations between the modes play a role. 

The total excess spectrum can be reproduced by adding the largest
terms of the cross-contributions between the modes, which are given
by

\begin{equation}
\delta\mathbf{C}^{\mathrm{cross}}(\omega)=\sum_{n,m=1,m<n}^{8}\Re\left(\frac{\beta_{\mathrm{ext},nm}(\omega)}{(1-\lambda_{n}(\omega))(1-\lambda_{m}^{*}(\omega))}\,\mathbf{u}_{n}(\omega)\mathbf{u}_{m}^{T*}(\omega)\right).\label{eq:spec_cross}
\end{equation}
The contributions to the spectrum due to the crosscorrelations between
the dominant modes are shown in \prettyref{fig:Composition-of-power}A3
and \prettyref{fig:Composition-of-power}B3. For population 2/3E the
crosscorrelation between the two modes is positive and provides the
missing offset to the excess spectrum. For population 2/3I, the crosscorrelation
between the mode associated to the low-$\gamma$ peak and the self-coupling
of population 2/3I is negative below $40\Hz$.

In other words, providing oscillatory input between $0\Hz$ and $40\Hz$
to population 2/3I elicits large responses of two modes. The responses
of the modes are given by weighted linear combinations of the population
rates corresponding to a multiplication of the rate vector containing
all rates with the left eigenvector of the modes. However, the anti-correlation
between the modes at low frequencies, which is strongest at $20\Hz$,
lowers the fluctuations in the total response of population 2/3I.
Since the amplitude of the negative correlation decreases for larger
frequencies, a peak appears in the power ratios. This peak could be
misinterpreted as the self-coupling of population 2/3I to generate
an oscillation in the low-$\gamma$ frequency range, even though the
peak is generated by a more complex sub-circuit in the upper layers.

\section*{Discussion}

In this manuscript we analyze the responses of networks, that generate
oscillations internally by means of embedded sub-circuits, to oscillatory
stimuli and demonstrate how the stimulus is reflected in different
measures of the network response. We first employ a sequence of phenomenological
rate models, each of which illustrating one dynamical effect that
crucially shapes the response. Finally, we apply these insight to
circuits composed of populations of LIF model neurons that can be
mapped to the former class of models analytically. Our results shed
light on the amount of information regarding the anatomical origin
of produced oscillations that can be inferred from the responses to
oscillatory stimuli. The insights outlined here can be exploited to
design stimulation protocols that uncover dynamically relevant sub-circuits.

Theoretical advances have been made to explain the results of Cardin
et al. \citep{Cardin2009}, who have shown that inhibitory neurons
exhibit a resonance in the $\gamma$ frequency regime when stimulated
with oscillatory light pulses, while excitatory neurons amplify low
frequencies. Theoretical studies reproduced these results by considering
certain neuronal or synaptic dynamics. Tchumatchenko et al. \citep{Tchumachenko2014}
assume sub-threshold resonances of inhibitory neurons as well as gap
junctions between them and Tiesinga \citep{Tiesinga12} considers
slow synaptic currents for the connections from the excitatory to
the inhibitory neurons. Here we investigate the effect of the network
architecture on the measured responses. Using three characteristic
connectivity motifs of an E-I-circuit, we identify connectivity motifs
that yield responses resembling those shown in Cardin et al. We also
demonstrate how these results crucially depend on the size of the
stimulus as well as on the measure of the network response.

\subsection*{The effect of small versus large stimulus amplitudes}

The effect of stimuli on the dynamics of a circuit can be classified
into three regimes. In the simplest case, additional input can be
treated as a small perturbation, which does not influence the dynamical
state of the circuit (as done in \citep{Tchumachenko2014}). In this
case, the stimulus can be treated analogously to the internally generated
noise of the population rates, which is, for example, produced within
networks of LIF neurons of finite size \citep{Brunel99,Lindner05_061919,Pernice11_e1002059,Trousdale12_e1002408,Pernice12_031916,Tetzlaff12_e1002596,Helias13_023002}.
In this setting, external input to populations of LIF neurons can
be described as a modulation of the synaptic input, which is filtered
by the dynamic transfer function of the populations the stimulus is
applied to. The dynamic transfer function is typically a low-pass
filter yielding the emphasis of low frequencies by the network response.
This explanation is similar to the results in \citep{Barbieri2014},
where the externally applied signal is described by an Ornstein-Uhlenbeck
process and the low-pass filter is therefore explicitly introduced.
Since the elevated low-frequency components in the network response
contain the filter properties of individual populations rather than
network dynamics, we suggest a modified stimulation protocol which
eliminates the effect of the initial filtering and therefore enables
the analysis of a signal which emerges from internal network dynamics
alone. 

An input that affects the stationary properties of the circuit may
change the working point within the linear regime of the static transfer
function. The dynamical properties of the circuit then stay approximately
unaltered and only the change of the stationary rates needs to be
accounted for. In contrast, an input that changes the stationary rates
in the nonlinear regime also changes the dynamic transfer functions
of the populations, which potentially alters the dynamic behavior
of the entire network. We show that this change can often be approximated
by adjusting the prefactor of the dynamic transfer function (see \nameref{subsec:Approximation_of_dtf}).
Applying this approximation to describe constant positive input to
a self-coupled inhibitory population, which generates a distinct frequency,
shows that the peak in the spectrum increases, while its peak frequency
remains unaltered. In other words, the eigenvalue trajectory which
determines the dynamics of the circuit is shifted towards instability
by the constant input, while its shape remains unaltered. Since this
effect dominates the response spectrum compared to changes evoked
by a purely oscillatory stimulus, this finding supports the statement
by Tiesinga et al. \citep{Tiesinga09}, who suggested to employ constant
stimuli as an alternative to oscillatory stimuli to investigate the
origin of oscillations experimentally. The result is in line with
the high sensitivity of the low-$\gamma$ power to the external input
to population 4E: the microcircuit model used in this paper is derived
from the model employed in \citep{Bos16_1} by reducing the external
input to population 4E and shows strongly reduced oscillations in
the low-$\gamma$ range.

The approximation of the modified dynamic transfer function by a change
in the prefactor needs to be applied with caution, when the population
is embedded in a network. Here the static transfer function can show
higher degrees of nonlinearity due to the recurrent feedback. As a
result a stimulus is more likely to change the dynamic structure of
the network. These effects can only be captured by linearizing the
system around the new working point. An analytical description of
the transition between the resting state and the stimulus induced
state remains to be investigated. Recent results on the nonlinear
transfer function of the LIF model will be useful in this approach
\citep{Voronenko2017}. 

In summary, the results derived here show that the way in which a
stimulus affects the dynamical regime of a network, as well as the
filtering properties of the populations, should be taken into account
when probing a network for the origin of internally generated oscillations.

\subsection*{Power ratios versus response spectra}

Experimental studies \citep{Cardin2009} demonstrate the responses
to oscillatory stimuli by means of power ratios, the LFP power at
stimulus frequency normalized by the LFP power at that frequency at
rest. Theoretical studies, on the contrary, consider normalized network
responses \citep{Tchumachenko2014} as well as absolute responses
at stimulus frequency \citep{Tiesinga12}.

We analyze the effectiveness of detecting the anatomical origin of
oscillations when using power ratios compared to response spectra
evoked by oscillatory stimuli. Power ratios can yield misleading results
if the spectrum at rest is not entirely flat. In these cases it appears
advantageous to consider the difference of the spectra with and without
additional stimulus. Small stimuli that do not affect the stationary
dynamics of the circuit evoke responses of oscillatory modes that
are also responsible for the oscillations in the resting condition.
The peaks may be canceled in the relative spectrum and therefore information
can get lost when considering power ratios. In particular, in one-population
systems the power ratio becomes independent of the intrinsic resonance
phenomena of the network and reflects the filter properties of the
population. In higher dimensional systems, a stimulation protocol
which allows for the reconstruction of all circuit internal variables
(\prettyref{eq:spectrum_decomp_with_input}) can, in principle, be
designed from the population rate spectra and cross-spectra obtained
by separately stimulating each population. Applications of stimuli
that affect the stationary dynamics of the circuit in the nonlinear
regime, however, change the dynamic response properties of the circuit.
If the response properties are changed such that the resonance of
an oscillatory mode is strengthened, these effects can dominate and
also show up in power ratios. Therefore, we propose to consider both,
absolute and relative spectra.

\subsection*{Connecting network responses to dynamic network architecture}

Oscillations induced by the network structure can either be generated
by the self-coupling of one population and imposed onto other populations
or by the interplay of several populations. We analyze here whether
the involvement of one population in the generation of an oscillation
can be investigated by means of oscillatory stimuli to that population.
We show how the network response to oscillatory stimuli can be decomposed
into responses generated by the auto- as well as crosscorrelation
of dynamic modes. Each mode can, individually, be traced back to its
anatomical origin, namely the sub-circuit that generates the associated
oscillation. However, in the analysis of experimental or simulated
data, such decomposition into modes is inaccessible. The anatomical
origin of the oscillation could therefore only be inferred from responses
generated by a single mode. It turns out that the stimulation of an
individual population elicits the response of a single mode only in
trivially connected circuits, in which the connections between populations
are negligible. In more complicated circuits, populations typically
participate in multiple dynamical modes, which are activated together
when stimulating that population.

The mode composition of the network response depends on the considered
measure of the network response. Experiments typically measure LFP
responses \citep{Cardin2009}, which have been linked to the input
onto excitatory neurons \citep{Linden11_859,Mazzoni2015}.  Tchumachenko
et al. \citep{Tchumachenko2014} defined the network response as the
response of the population that is stimulated and therefore measured
different responses depending on the stimulus. Tiesinga \citep{Tiesinga12}
considers the activity of the excitatory cells. These two theoretical
studies referred to the findings presented by Cardin et al. who showed
that that the $\gamma$-resonance was present in the LFP ratio when
stimulating the inhibitory cells at $\gamma$ frequency, but was missing
when stimulating the excitatory cells. Given that the connectivity
plays a role in the composition of the LFP, it is possible that a
resonance is visible in the population spectrum, but not in the LFP
response (see for example the spectra of the two-dimensional network
without self-coupling \prettyref{fig:2d_PING}B). In the presented
example, the feedback of the excitatory population response is missing
and the $\gamma$ rhythm is therefore not relayed back to the pyramidal
neurons where it would contribute to the LFP. Even though an E-I network
with a missing E-E-loop might not be biologically realistic, the same
effect could be caused by a large amount of NMDA receptors at the
synapses of the excitatory neurons: The slow synapses then act as
a low-pass filter which the $\gamma$ oscillation cannot pass (the
same mechanism was investigated in \citep{Tiesinga12}). In other
words, the connection would not be present dynamically at $\gamma$
frequency.

To test the hypothesis that the findings of Cardin et al. suggest
an oscillation generating mechanism which solely involves the inhibitory
neurons, we compare the responses of two exemplary circuits (see \nameref{subsec:ING1}
and \nameref{subsec:ING2}). In the first circuit, the $\gamma$ oscillation
is generated by the I-I-loop and subsequently imposed onto the excitatory
population. The second circuit, in contrast, requires all connections
for the generation of $\gamma$. We show that the LFP response to
oscillatory stimulation of the inhibitory neurons shows a resonance
at $\gamma$, while the response to stimulated excitatory neurons
resembles a low pass filter, regardless of the origin of the oscillation.
We therefore conclude, similarly to \citep{Tiesinga09}, that oscillatory
stimuli cannot exclude the involvement of excitatory neurons in the
oscillation generating mechanism.

We discuss the design of a stimulation protocol that isolates the
responses of individual dynamic modes by exciting populations in the
same ratio in which they contribute to the oscillation generating
sub-circuit. However, it remains an open question how these single
mode responses can be distinguished from mixtures of mode responses
without the knowledge of the dynamic transfer of the mode. It is also
debatable whether this protocol is experimentally feasible, given
that, if the structure of the mode were unknown, numerous runs in
which several populations are stimulated with various strength and
time lags would be required. 

\subsection*{The emergence of ambiguous resonances in the network response of
a microcircuit model}

Applying oscillatory stimuli to the populations in a multi-layered
model of a column composed of LIF model neurons demonstrates that
the response spectra in large spiking networks can be predicted theoretically.
The results (\prettyref{fig:Power-ratios-sim}) show that stimuli
evoking firing rate fluctuations as large as the firing rate itself
(see left panel in \prettyref{fig:Population-rate-responses}C) as
well as response spectra of amplitudes comparable to those evoked
in experiments \citep{Cardin2009} (see right panels in \prettyref{fig:Population-rate-responses}),
can still be sufficiently well described by the employed theoretical
framework, which is based on mean-field and linear response theory.

The power ratios of population rates with oscillatory stimuli applied
to the respective population reveal a resonance in the low-$\gamma$
frequency range when stimulating population 2/3I. However, it has
been shown that the low-$\gamma$ peak is generated within a sub-circuit
which is located in the upper layers and involves several populations
\citep{Bos16_1}. Decomposing the network response at $\gamma$ frequency
into contributions of the dynamic modes that shape the oscillations
in the microcircuit model, reveals that the response is mainly shaped
by two modes including their crosscorrelation; in addition to the
mode that is responsible for the low-$\gamma$ peak, population 2/3I
contributs strongly to the mode which is composed of the 2/3I-2/3I
loop and which is responsible for the generation of a peak at very
high frequencies \citep{Bos16_1}. Stimulating population 2/3I therefore
elicits responses of both modes, which are anti-correlated for low
frequencies and therefore cancel the contributions of the autocorrelations
of the modes. This cancellation for low frequencies, but not for high
frequencies, gives rise to a peak in the power ratio that could be
misinterpreted as the signature of an underlying I-I loop that generates
the low-$\gamma$ peak.

In summary, we demonstrate the importance of correctly identifying
the dynamic influence of the stimulus on the system as well as the
considered output measure when interpreting experimental results.
By analyzing reduced circuits as well as a model of a column in the
primary sensory area, we demonstrate that the entire underlying network
needs to be taken into account when interpreting emerging signals
with respect to the origin of oscillations.

\section*{Methods\label{sec:Methods}}

\subsection*{Static and dynamic transfer function of LIF-neurons\label{subsec:Static-and-dynamic}}

The description of the population dynamics discussed here follows
the outline in \citep{Ostojic11_e1001056} and the terminology has
been introduced in \citep{Ledoux11_1}. Activity entering one population
can be regarded as first passing a linear filter $g(t)$ (dynamic
transfer function) and subsequently being sent through a static nonlinear
function (static transfer function) (see Fig. 1B in \citep{Ostojic11_e1001056}).
Hence the rate of one population of unconnected neurons receiving
white noise input $x(t)$ with strength $I_{0}$ can be described
as

\begin{equation}
r(t)=\nu([g\ast I_{0}x](t))\approx r_{0}+I_{0}\nu'(0)\,[g\ast x](t),\label{eq:LN-model}
\end{equation}
where $\ast$ denotes the convolution of two signals. In the second
step, the nonlinear function was linearized around the static point
(also referred to as the working point) with $\nu(0)=r_{0}$. The
linearized version of the linear-nonlinear model above can be mapped
to the dynamics of LIF neuron models. Here, we consider LIF model
neurons with exponentially decaying synaptic currents, i.e. with synaptic
filtering. The dynamics of the membrane potential $V$ and synaptic
current $I_{\mathrm{s}}$ are given by \citep{Fourcaud02}
\begin{eqnarray}
\taum\frac{dV}{dt} & = & -V+I_{s}(t)\nonumber \\
\taus\frac{dI_{s}}{dt} & = & -I_{s}+\taum\sum_{j=1}^{N}J_{j}\,s_{j}(t-d_{j})\,,\label{eq:diffeq_iaf}
\end{eqnarray}
where $\taum$ is the membrane time constant and $\taus$ the synaptic
time constant. The membrane resistance $\taum/C_{\mathrm{m}}$ has
been absorbed into the definition of the current. Input is provided
by the presynaptic spike trains $s_{j}(t)=\sum_{k}\delta(t-t_{k}^{j})$,
where the $t_{k}^{j}$ mark the time points at which neuron $j$ emits
an action potential. The synaptic efficacy is denoted as $J_{j}=\taus w_{j}/C_{m}$,
with $w$ in Ampere. Whenever the membrane potential $V$ crosses
the threshold $V_{\theta},$ the neuron emits a spike and $V$ is
reset to the potential $V_{\mathrm{r}}$, where it is clamped during
$\taur$. In the diffusion approximation the dynamics reads \citep{Fourcaud02}
\begin{eqnarray}
\taum\frac{dV}{dt} & = & -V+I_{\mathrm{s}}(t)\nonumber \\
\taus\frac{dI_{\mathrm{s}}}{dt} & = & -I_{\mathrm{s}}+\mu+\sigma\sqrt{\taum}\,\xi(t),\label{eq:diffeq_iaf-1-1-1-1}
\end{eqnarray}
where the input to the neuron is characterized by its mean $\mu$
and a variance proportional to $\sigma$, and $\xi$ is a centered
Gaussian white process satisfying $\langle\xi(t)\rangle=0$ and $\langle\xi(t)\xi(t^{\prime})\rangle=\delta(t-t^{\prime})$.
The static transfer function $\nu$ can be obtained for white noise
(originating from $\delta$-synapses, i.e. $\taus=0$) \citep{Siegert51}
or colored noise (originating from filtered synapses, i.e. $\taus\ne0$)
\citep{Fourcaud02}. The stationary rate is then given by $r_{0}=\nu(\mu,\sigma)$.
The dynamic transfer function $h(t)=\nu'(0)g(t)$ has been derived
in the Fourier domain using linear response theory to systems exposed
to white \citep{Brunel99} and colored noise \citep{Schuecker15_transferfunction}.
To employ linear response theory, the system has to be linearized
around the static point, yielding a dynamic transfer function that
also depends on the working point $H(\omega):=H(\omega,\mu,\sigma)$.
The dynamic transfer function of the LIF model with $\delta$-synapses
is given by \citep{Brunel99}

\begin{equation}
H_{\mathrm{WN}}(\omega,\mu,\sigma)=G\,\frac{1}{1+i\omega\taum}\frac{\Phi_{\omega}^{\prime}\vert_{x_{\theta}}^{x_{R}}}{\Phi_{\omega}\vert_{x_{\theta}}^{x_{R}}},\label{eq:transfer_function_white_noise-1}
\end{equation}
where $G=\frac{\sqrt{2}r_{0}}{\sigma}$ and we introduced $\Phi_{\omega}(x)=u^{-1}(x)\,U(i\omega\tau-\frac{1}{2},x)$
as well as $\Phi_{\omega}^{\prime}=\partial_{x}\Phi_{\omega}$. Here,
$U(i\omega\tau-\frac{1}{2},x)$ is the parabolic cylinder function
\citep{Abramowitz74,Lindner01_2934} and the boundaries are $x_{\{R,\theta\}}=\sqrt{2}\,\frac{\{V_{R},V_{\theta}\}-\mu}{\sigma}$.
The effect of the synaptic filtering $\sim\taus$ is twofold: First,
input is low-pass filtered by the factor $\frac{1}{1+i\omega\taus}$
appearing in the transfer function. Second, it causes a shift of the
boundaries \citep{Schuecker15_transferfunction}, i.e. $x_{\{\tilde{R},\tilde{\theta}\}}=\sqrt{2}\frac{\{V_{R},V_{\theta}\}-\mu}{\sigma}+\sqrt{\frac{\taus}{2\taum}}$,
which is correct up to linear order in $k=\sqrt{\frac{\taus}{\taum}}$
and valid up to moderate frequencies. Finally the dynamical transfer
function is given by 
\begin{equation}
H(\omega,\mu,\sigma)=G\,\frac{1}{1+i\omega\taum}\frac{1}{1+i\omega\taus}\frac{\Phi_{\omega}^{\prime}\vert_{x_{\tilde{\theta}}}^{x_{\tilde{R}}}}{\Phi_{\omega}\vert_{x_{\tilde{\theta}}}^{x_{\tilde{R}}}}.\label{eq:transfer_function_colored_Noise-1}
\end{equation}
Note that we only consider the dominant part of the dynamical transfer
function, i.e. the modulation of the output rate caused by a modulation
of the mean input. The part of the transfer function corresponding
to a modulation of the variance of the neurons' input \citep{Lindner01_2934,Schuecker15_transferfunction}
is one order of magnitude smaller and neglected here. The formalism
for rate fluctuations of a single unconnected population can be extended
to an $N$-dimensional recurrent network of populations with the connectivity
matrix $\mathbf{M}^{A}$ and delays $d$, where each population receives
input from other populations, each of which described as a rate $r_{i}(t)$
with additional noise $x_{i}(t)$ which is subsequently filtered by
the population specific dynamic transfer function $h_{i}(t)$ 

\begin{equation}
r_{i}(t)=h_{i}\ast\sum_{j=1}^{N}M_{ij}^{A}\left(r_{j}(\circ-d_{ij})+x_{j}(\circ-d_{ij})\right).\label{eq:rate_convolution}
\end{equation}
Here, $h$ is obtained from the Fourier transform of $H$. The connectivity
matrix $\mathbf{M}^{A}$ follows from the LIF-network parameters,
i.e. $M_{ij}^{\mathrm{A}}\equiv\taum I_{ij}J_{ij}$, with $I_{ij}$
being the indegree from population $j$ on population $i$.

\subsection*{Approximation of the dynamic transfer function\label{subsec:Approximation_of_dtf}}

The dynamical response to a constant current, in the following termed
the DC limit, can be obtained by evaluating the dynamic transfer function
at frequency zero, i.e.
\begin{eqnarray}
A(\mu,\sigma):=H(0,\mu,\sigma) & = & \frac{\partial\nu(\mu,\sigma)}{\partial\mu}\,.\label{eq:def_DClim}
\end{eqnarray}
The equal sign follows from the fact that the integral over the impulse
response $h(t)=\iFtr{H(\omega)}t$ is given by the response to a constant
input \citep{Oppenheim96}. We now investigate how the dynamic transfer
function behaves for an isolated population which receives external
input defined by its mean $\mu$ and variance $\propto\sigma^{2}$

\begin{align}
\mu & =\taum K_{\mathrm{ext}}J_{\mathrm{ext}}\nu_{ext},\qquad\sigma^{2}=\taum K_{\mathrm{ext}}J_{\mathrm{ext}}^{2}\nu_{ext}.\label{eq:working_point}
\end{align}
Here $K_{\mathrm{ext}}$ denotes the external number of synapses weighted
by $J_{\mathrm{ext}}$, with the external firing rate $\nu_{ext}$.
Perturbing the external rate typically yields a larger change of the
mean than the variance ($\frac{d\mu}{d\nu_{\mathrm{ext}}}\propto K_{\mathbb{\mathrm{ext}}}$
,$\frac{d\sigma}{d\nu_{\mathrm{ext}}}\propto\sqrt{K_{\mathrm{ext}}}$,
with $K_{\mathrm{ext}}\approx10^{3}$).

\begin{figure}[ht]
\begin{centering}
\includegraphics[scale=0.8]{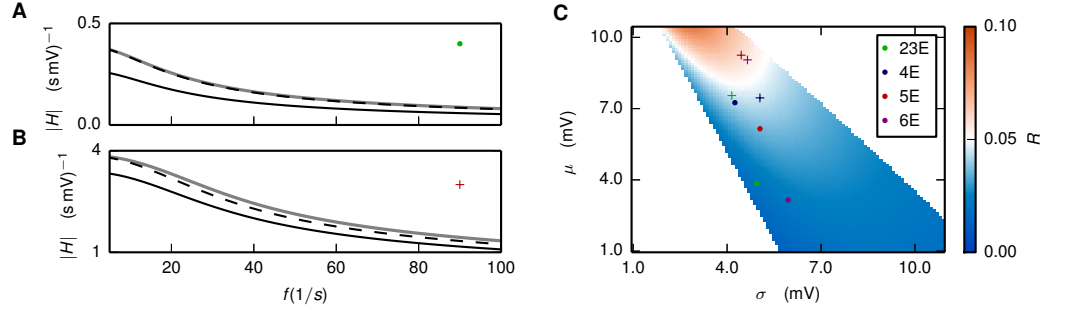}
\par\end{centering}
\caption{\label{fig:transfer_function_lif}\textbf{A} Absolute value of dynamic
transfer function $H$ for $(\mu,\sigma)_{2/3E}=(3.9,5.0)\protect\mV$
(black), $(\mu,\sigma)=(4.4,5.0)\protect\mV$ (gray) and $H^{approx}$
(black, dashed). Other parameters are $\protect\taum=10\protect\ms$,
$\protect\taus=0.5\protect\ms$, $V_{\theta}=-55\protect\mV$ and
$V_{R}=-70\protect\mV$. Parameters correspond to population 2/3E
(green cross in \textbf{C}). \textbf{B} Absolute value of dynamical
transfer function $H$ for $(\mu,\sigma)_{5I}=(9.3,4.5)\protect\mV$
(black), (gray) $(\mu,\sigma)=(9.8,4.5)\protect\mV$ and $H^{approx}$(black,
dashed). Parameters correspond to population 5I (red cross in \textbf{C}).\textbf{
C }Relative error $R(\mu,\sigma,\mu^{\prime})$ of the approximation
given by \prettyref{eq:rel_error} for $\delta\mu=0.5\protect\mV,\,\omega_{max}=2\pi\cdot200\protect\Hz$.
Choice of $\delta\mu$ corresponds to a relative change of the rate
$r_{0}$ of $25$ percent on average. Qualitatively similar results
are obtained for different values of $\delta\mu$, where the error
$R$, in general, increases with increasing $\delta\mu$.  Values
of $\mu$ and $\sigma$ are constrained to regions in which the resulting
stationary rate $r_{0}$ is in the range $r_{0}\in(0.1\protect\Hz,10\protect\Hz)$.
Since the static transfer function is strictly increasing with $\sigma$
the left edge of the shown region corresponds to $r_{0}=0.1\protect\Hz$
and the right edge to $r_{0}=10\protect\Hz$. The working points for
the $8$ populations of the microcircuit are marked with symbols (see
legend, crosses mark respective inhibitory populations).}
\end{figure}

We therefore neglect the variation
of $\sigma$ and restrict the analysis to a perturbation of the mean,
i.e. $\mu\rightarrow\mu^{\prime}=\mu+\delta\mu$. \prettyref{fig:transfer_function_lif}A
shows that the DC limit of $H$ significantly changes while its shape
stays approximately constant. This suggests that we can approximate
$H(\omega,\mu^{\prime},\sigma)$ by a change of the DC-limit by altering
the prefactor in the following way 
\begin{eqnarray}
H(\omega,\mu^{\prime},\sigma) & \approx & \frac{A(\mu^{\prime},\sigma)}{A(\mu,\sigma)}\,H(\omega,\mu,\sigma)=:H^{approx}(\omega,\mu,\mu^{\prime},\sigma).\label{eq:H_approx}
\end{eqnarray}
In the approximation above a change in the input yields the following
dynamic transfer function

\begin{align}
H_{I}(\omega) & =H(\omega,\mu',\sigma)=\left(1+\frac{\delta A(\mu,\mu',\sigma)}{A(\mu,\sigma)}\right)H(\omega,\mu,\sigma)\nonumber \\
 & \mbox{with}\quad\delta A(\mu,\mu',\sigma)=\frac{\partial^{2}\nu}{\partial\mu^{2}}\,\delta\mu.\label{eq:H_I}
\end{align}

This approximation can be evaluated defining the relative error
\begin{eqnarray}
R(\mu,\sigma,\mu^{\prime}) & = & \frac{\int_{0}^{\omega_{max}}(|H^{approx}(\omega,\mu,\mu^{\prime},\sigma)|-|H(\omega,\mu^{\prime},\sigma)|)d\omega}{\int_{0}^{\omega_{max}}(|H(\omega,\mu^{\prime},\sigma)|)d\omega},\label{eq:rel_error}
\end{eqnarray}
which is shown in \prettyref{fig:transfer_function_lif}C. In the
fluctuation driven regime, which corresponds to low values of $\mu$
and high values of $\sigma$ (bottom part of the figure), $H^{approx}$
constitutes a good approximation. In the regime with large $\mu$
and low $\sigma$ the approximation $H^{approx}$ is less accurate
since the change in the shape of $H$ is not negligible (\prettyref{fig:transfer_function_lif}B),
in line with the finding of a resonance at the firing rate in the
mean driven regime \citep{Lindner01_2934}. 

In conclusion, in the fluctuation driven regime the perturbation can
be approximately absorbed into the prefactor of the dynamic transfer
function. Note that the DC-limit does not change for variations of
$\mu$ that affect the static transfer function $\nu(\mu,\sigma)$
in the linear regime, where $\partial\nu(\mu,\sigma)/\partial\mu$
is constant by definition. However, it turns out that the static transfer
functions of the populations with working points equal to those in
the microcircuit model (shown in \ref{fig:transfer_function_lif}C)
are affected nonlinearly by a perturbation in $\mu$ ($\partial^{2}\nu(\mu,\sigma)/\partial\mu^{2}\neq0$).

\begin{figure}[h]
\centering{}\includegraphics[scale=0.8]{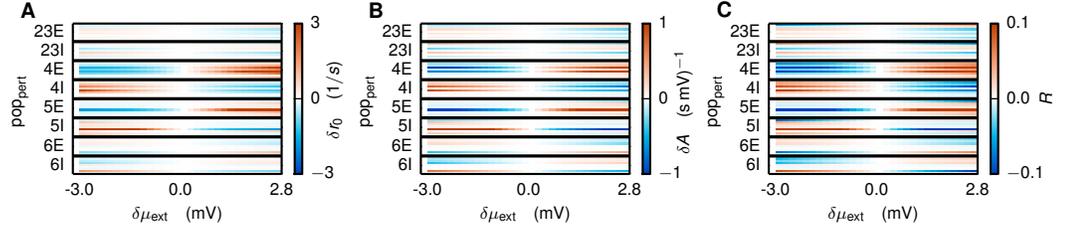}\caption{\label{fig:Perturbation-of-microcircuit}\textbf{Perturbation of the
input to the microcircuit}. \textbf{A }One block of eight rows shows
the change in the stationary rate $\delta r_{0}$ of the populations
in the microcircuit in response to a perturbation $\delta\mu_{\mathrm{ext}}$
to population $\mathrm{pop}_{\mathrm{pert}}$, as specified on the
y-axis. Each individual block is ordered from top to bottom according
to the populations {[}2/3E, 2/3I, 4E, 4I, 5E, 5I, 6E, 6I{]} with black
rows separating the blocks. For example, the second line in the top
row shows the change in the stationary rate of population 2/3I after
perturbing the external input to population 2/3E. The range of $\delta\mu_{\mathrm{ext}}$
corresponds roughly to a change of the external rate ($r_{\mathrm{ext}}=8\protect\Hz)$
of about $\pm1\protect\Hz$ \textbf{B }Changes of the DC-limit to
perturbation. Structure as in \textbf{A}. \textbf{C: }Changes of the
error $R$ to perturbation. Structure as in \textbf{A}.}
\end{figure}

So far, populations were considered in isolation. To investigate how
a perturbation effects the dynamic transfer function of a population
embedded into a network, we now treat a perturbation of the mean external
input $\mu_{\mathrm{ext}}$ to a population in the microcircuit model.
Since the stationary rates of the populations in the network depend
on each other, the static transfer function needs to be solved self-consistently
when introducing a perturbation to one population. This yields a new
stationary rate $r_{0}^{\prime}=r_{0}+\delta r_{0}$ and working point
for each population. We first investigate the induced changes in the
rates (\prettyref{fig:Perturbation-of-microcircuit}A). A perturbation
to one population has an effect on all eight populations, where by
trend a positive perturbation to an excitatory population causes an
increase in the rates while the opposite is true for a perturbation
of an inhibitory population (compare $\mathrm{pop_{\mathrm{pert}}}=\mathrm{4E}/4I$).
However, increased input can also yield higher rates in some populations
and lower ones in others (see $\mathrm{pop_{\mathrm{pert}}}=6E$ and
$\mathrm{pop_{\mathrm{pert}}}=6I$). Another tendency is that excitatory
populations are more strongly affected by perturbations than inhibitory
ones. In particular, population 5E is very sensitive to perturbations
of populations in L4 or L5, while populations in L4 are very sensitive
to perturbations in L4.

We further investigate the corresponding changes in the DC-limit of
the dynamic transfer function (\prettyref{fig:Perturbation-of-microcircuit}B).
In general the DC-limit follows the changes of the rates. However,
some differences can be observed: for example when perturbing population
4E the rate of population 5I is sensitive to the perturbation, but
the DC-limit stays almost constant, which hints on the perturbation
acting on the linear regime of the static transfer function of population
5I. In summary, comparing the response of a population in isolation
and embedded in a network to a perturbation in its input shows that
the network structure can amplify or decrease as well as reverse the
sign of the response. The responses of the other populations to this
perturbation can be uniform as well as diverse.

The relative error, which here needs to account for both, the change
in $\mu$ as well as the change in $\sigma$ induced by rate changes
of the other populations (compared to \prettyref{eq:rel_error} which
does not depend on changes in $\sigma$), reads 
\begin{eqnarray*}
R(\mu,\sigma,\mu^{\prime}) & = & \frac{\int_{0}^{\omega_{max}}(|H^{approx}(\omega,\mu,\mu^{\prime},\sigma,\sigma^{\prime})|-|H(\omega,\mu^{\prime},\sigma^{\prime})|)d\omega}{\int_{0}^{\omega_{max}}(|H(\omega,\mu^{\prime},\sigma^{\prime})|)d\omega}
\end{eqnarray*}
and is shown in \prettyref{fig:Perturbation-of-microcircuit}C. The
error follows the behavior of the rate and the DC-limit and therefore
shows that the higher the changes in the working point of the populations
the higher the error. The error of the approximation of the dynamic
transfer functions is large compared to the error for the same populations
in isolation. However, it stays within the limits of $10\%$ given
an alteration of the input of the same order. How these changes effect
the prediction of the spectrum remains to be investigated.

\subsection*{Mapping changes in the stationary rate to changes in the eigenvalue\label{subsec:params_toy_models}}

We identify the eigenvalue trajectory of the one-dimensional circuit
(discussed in \nameref{subsec:I-I-loop}) as the weighted dynamic
transfer function

\begin{equation}
\lambda(\omega)=-wH(\omega),\label{eq:eigenvalue_trajectory}
\end{equation}
where $\lambda(\omega)$ denotes the Fourier transformation of the
time dependent eigenvalue defined as $\tilde{\lambda}(t)=\frac{1}{2\pi}\int\lambda(\omega)\,e^{i\omega t}\,d\omega$.
The eigenvalue trajectory of the circuit with an additional large
constant input reads

\begin{equation}
\lambda_{I}(\omega)=-wH_{I}(\omega)\approx-w(1+\delta A/A)H(\omega),\label{eq:eigenvalue_trajectory_input}
\end{equation}
where we inserted the approximation of the dynamical transfer function,
which is discussed in the previous section. Changes in the eigenvalue
can therefore be parameterized as 
\[
\lambda_{I}=(1+\alpha_{\lambda})\lambda,
\]
with $\alpha_{\lambda}=\delta A/A$ being the ratio by which the
prefactor of the dynamic transfer function is shifted and which is
related to the excitability of the circuit (\prettyref{fig:eigenvalue_trajectory}).
The frequency dependence of the eigenvalues was omitted for clarity
of notation. The following considerations show how the shift in the
eigenvalue relates to the change of the stationary rate.

A constant stimulus is applied by an increase in the external rate
($\tilde{\nu}_{\mathrm{ext}}=(1+\alpha_{\mu})\nu_{\mathrm{ext}}$),
yielding a change in the mean value of the input current ($\delta\mu_{\mathrm{ext}}=\alpha_{\mu}\mu_{ext}$,
see \prettyref{eq:working_point}). Following the argument in the
previous section, we neglect changes in the variance. The change in
the external input yields the following change in the stationary rate
$r_{0}$

\begin{equation}
\alpha_{r}r_{0}=\left.\frac{\partial\nu}{\partial\mu}\right|_{\mu=\mu_{\mathrm{ext}}}\delta\mu+\frac{1}{2}\left.\frac{\partial^{2}\nu}{\partial\mu^{2}}\right|_{\mu=\mu_{\mathrm{ext}}}\delta\mu^{2},\label{eq:rate_change}
\end{equation}
where higher orders in the derivative of the static transfer function
$\nu(\mu,\sigma)$ were neglected. In this recurrent network $\delta\mu$
is composed of the perturbation in the external input $\delta\mu_{\mathrm{ext}}$
in addition to a contribution from the feedback connection $\delta\mu_{\mathrm{\mathrm{rec}}}=-wr_{0}$.
We now identify that $A=\left.\frac{\partial\nu}{\partial\mu}\right|_{\mu=\mu_{\mathrm{ext}}}$,
abbreviate $m=\left.\frac{\partial^{2}\nu}{\partial\mu^{2}}\right|_{\mu=\mu_{\mathrm{ext}}}$and
recall that $\delta A=m\delta\mu$ (\prettyref{eq:H_I}). Inserting
this in the equation above yields the following relation between the
ratio of the eigenvalue shift and the rate change

\begin{equation}
\frac{\left(\alpha_{\lambda}+1\right)^{2}-1}{2\alpha_{r}}=\frac{r_{0}m}{A^{2}}.\label{eq:dlambda_dr}
\end{equation}
This shows that in this approximation the eigenvalue does not shift,
if the working point sets the static transfer function in the linear
regime, i.e. $m=0$. However, we demonstrated that, in particular,
in recurrent networks the nonlinear effect can play a role (\prettyref{fig:Perturbation-of-microcircuit}).

For the self-coupled inhibitory units (discussed in \nameref{subsec:I-I-loop}),
we chose the following parameters:
\begin{align}
\alpha_{\lambda} & =0.15\:,r_{0}/N=1\Hz,\:\alpha_{r}=0.8,\:w=-4\:mV\nonumber \\
 & \mathrm{\mbox{yielding}}\quad A=0.05\:\mathrm{mV}^{-1},\:A_{I}=0.058\:\mathrm{mV}^{-1}\nonumber \\
 & \mbox{and}\quad\tau=2\ms,\:d=3.6\ms,\:I_{0}=0.5/\pi.\label{eq:params_1d_model}
\end{align}
The parameters of the dynamic transfer function for both populations
in the two-dimensional circuits (discussed in \nameref{subsec:2d})
are given by:
\begin{align}
r_{0,\E}/N_{\E} & =r_{0,I}/N_{\I}=1\Hz,\:A=0.5\:\mathrm{mV}^{-1},\:w=1\mV\nonumber \\
\tau & =2\ms,\:d=1.5\ms,\:I_{0}=0.5/\pi.\label{eq:params_2d_model}
\end{align}

\subsection*{Composition of the spectrum \label{subsec:Composition-spec}}

The systems considered in this work are given by, or can be reduced
to, $N$-dimensional rate models with noise, while $N$ denotes the
number of populations. The spectrum of the populations is hence identical
to the diagonal of

\begin{equation}
\mathbf{C}(\omega)=\langle\mathbf{Y}(\omega)\mathbf{Y}^{\mathrm{T}}(-\omega)\rangle,\label{eq:spec0}
\end{equation}
where $\mathbf{Y}(\omega)$ is the rate vector in Fourier space, composed
of the rate $\mathbf{R}(\omega)$ and a noise term $\mathbf{X}(\omega)$
as $\mathbf{Y}(\omega)=\mathbf{R}(\omega)+\mathbf{X}(\omega)$. Inserting
the effective connectivity matrix, which is composed of the connection
strengths $M_{ij}$ and the dynamical transfer function of the populations
incorporating the connection delays $H_{i}(\omega)e^{-i\omega d_{ij}}$(for
further detail see \citep{Bos16_1}) into the self-consistent equation
for the population rates (\prettyref{eq:rate_convolution}), yields

\begin{equation}
\mathbf{C}(\omega)=(\mathbb{I}-\tilde{\mathbf{M}}_{\mathrm{d}}(\omega))^{-1}\,\mathbf{D}\,(\mathbb{I}-\tilde{\mathbf{M}}_{\mathrm{d}}(-\omega))^{-1,\mathrm{T}},\label{eq:spectrum}
\end{equation}
with the diagonal matrix $\mathbf{D}=\langle\mathbf{X}(\omega)\mathbf{X}^{\mathrm{T}}(-\omega)\rangle$
describing the power spectrum of the noise. In the reduction of spiking
networks, the noise is a finite-size effect with the autocovariance
$D_{ii}=\bar{r}_{i}/N_{i}\equiv r_{i}$ , where $\bar{r}_{i}$ is
the stationary firing rate of the $i$-th population and $N_{i}$
the number of neurons within population $i$.

The eigenvectors and eigenvalues of the effective connectivity matrix
are defined as

\begin{align}
\tilde{\mathbf{M}}_{d}(\omega)\mathbf{u}_{i}(\omega) & =\lambda_{i}(\omega)\mathbf{u}_{i}(\omega)\nonumber \\
\mathbf{v}_{i}^{T}(\omega)\tilde{\mathbf{M}}_{d}(\omega) & =\lambda_{i}(\omega)\mathbf{v}_{i}^{\mathrm{T}}(\omega).\label{eq:eigs_Meff}
\end{align}
The eigenvectors $\mathbf{u}_{i}(\omega)$ and $\mathbf{v}_{i}(\omega)$
with the convention $|\mathbf{u}_{i}(\omega)|^{2}=1$ and $\mathbf{v}_{i}^{\mathrm{T}}(\omega)\mathbf{u}_{j}(\omega)=\delta_{ij}$
constitute a bi-orthogonal basis. The eigenvectors and eigenvalues
of the term $(\mathbb{I}-\tilde{\mathbf{M}}_{\mathrm{d}}(\omega))^{-1}$
appearing in \prettyref{eq:spectrum} are given by

\begin{align}
(\mathbb{I}-\tilde{\mathbf{M}}_{\mathrm{d}}(\omega))^{-1}\mathbf{u}_{i}(\omega) & =\frac{1}{1-\lambda_{i}(\omega)}\mathbf{u}_{i}(\omega)\nonumber \\
\mathbf{v}_{i}^{\mathrm{T}}(\omega)(\mathbb{I}-\tilde{\mathbf{M}}_{\mathrm{d}}(\omega))^{-1} & =\frac{1}{1-\lambda_{i}(\omega)}\mathbf{v}_{i}^{\mathrm{T}}(\omega).\label{eq:decomp_propagator}
\end{align}
The diagonal noise correlation matrix $\mathbf{D}$ can be written
as the sum of outer products of the unit vectors $\mathbf{e}_{i}$,
which have the entry one at position $i$ and zero everywhere else,
weighted by the diagonal entries $r_{i}$

\begin{equation}
\mathbf{D}=\sum_{i}r_{i}\mathbf{e}_{i}\mathbf{e}_{i}^{\mathrm{T}}.\label{eq:D_old_basis}
\end{equation}
The unit vectors can be rewritten in the basis spanned by the eigenvectors
of the effective connectivity matrix

\begin{equation}
\mathbf{e}_{i}=\sum_{j}\alpha_{j}^{i}(\omega)\mathbf{u}_{j}(\omega)\,\mbox{with}\:\alpha_{j}^{i}(\omega)=\mathbf{v}_{j}^{\mathrm{T}}(\omega)\mathbf{e}_{i}.\label{eq:unit_vector_decomp}
\end{equation}
Here $\alpha_{j}^{i}(\omega)$ describes the projection of the $i$-th
unit vector onto the $j$-th eigenmode. Inserting the decomposition
of the unit vectors (\prettyref{eq:unit_vector_decomp}) into \prettyref{eq:D_old_basis}
yields the noise correlation matrix in the new basis

\begin{equation}
\mathbf{D}=\sum_{n,m}\beta_{nm}(\omega)\mathbf{u}_{n}(\omega)\mathbf{u}_{m}^{\mathrm{T}*}(\omega),\label{eq:D_new_basis}
\end{equation}
with $\beta_{nm}(\omega)=\sum_{i}r_{i}\alpha_{n}^{i}(\omega)\alpha_{m}^{i*}(\omega)$
weighing the $i$-th component of the $n$-th and $m$-th left eigenvector
with the rate of the $i$-th population. In other words, the $i$-th
population has a large contribution to $\beta_{nm}(\omega)$ if the
unit vector of population $i$ points in a similar direction as the
right eigenvector of the $n$-th and $m$-th mode and population $i$
has a large rate $r_{i}$. Here we also used that the effective connectivity
matrix is real valued in the time domain and therefore has the property
$\tilde{M}(-\omega)=\tilde{M}(\omega)^{\ast}$ in Fourier domain.
Inserting \prettyref{eq:D_new_basis} and the effective connectivity
matrix in the new basis (\prettyref{eq:decomp_propagator}) into the
expression for the spectrum (\prettyref{eq:spectrum}) results in 

\begin{align}
\mathbf{C}(\omega) & =\Big(\sum_{n}\frac{1}{1-\lambda_{n}(\omega)}\mathbf{u}_{n}(\omega)\mathbf{v}_{n}^{T}(\omega)\Big)\Big(\sum_{j,k}\beta_{jk}(\omega)\mathbf{u}_{j}(\omega)\mathbf{u}_{k}^{T*}(\omega)\Big)\nonumber \\
 & \quad\:\Big(\sum_{m}\frac{1}{1-\lambda_{m}^{*}(\omega)}\mathbf{v}_{m}^{*}(\omega)\mathbf{u}_{m}^{T*}(\omega)\Big)\nonumber \\
 & =\sum_{n,m}\frac{\beta_{nm}(\omega)}{(1-\lambda_{n}(\omega))(1-\lambda_{m}^{*}(\omega))}\mathbf{u}_{n}(\omega)\mathbf{u}_{m}^{T*}(\omega).\label{eq:spectrum_decomp}
\end{align}
The spectrum of the eight dimensional microcircuit is hence composed
of $64$ contributions. Eight contributions arise from individual
modes ($n=m$) and the remaining terms arise from contributions of
pairs of different modes ($n\neq m$).

\subsection*{Composition of the spectrum with input\label{subsec:spec_input}}

So far the contribution of external input on the spectrum produced
by the circuit has been neglected. In the microcircuit, all model
neurons receive external Poisson input from a population-specific
number of independent sources that each fire with the rate $\nu_{\mathrm{ext}}=8\Hz$.
Given that all $N_{\mathrm{ext},i}$ external sources are independent,
the total Poisson input received by the $j$th neuron in the $i$th
population can be approximated by Gaussian white noise described by 

\begin{align}
y_{\mathrm{ext},i}^{j}(t) & =\sum_{k=1}^{N_{\mathrm{ext},i}}\left(\nu_{\mathrm{ext}}+\sqrt{\nu_{\mathrm{ext}}}\chi_{\mathrm{ext},i}^{jk}(t)\right)=N_{\mathrm{ext},i}\nu_{\mathrm{ext}}+\sum_{k=1}^{N_{\mathrm{ext},i}}\sqrt{\nu_{\mathrm{ext}}}\chi_{\mathrm{ext},i}^{jk}(t)\nonumber \\
 & =N_{\mathrm{ext},i}\nu_{\mathrm{ext}}+\sqrt{N_{\mathrm{ext},i}\nu_{\mathrm{ext}}}\chi_{\mathrm{ext},i}^{j}(t).\label{eq:ExtInputToNeuron}
\end{align}
Here $\chi_{\mathrm{ext},i}^{jk}(t)$ is a random variable with zero
mean ($\langle\chi_{\mathrm{ext},i}^{jk}(t)\rangle=0$) and variance
$\langle\chi_{\mathrm{ext},i}^{jk}(t)\chi_{\mathrm{ext},i'}^{j'k'}(t')\rangle=\delta_{ii^{\prime}}\,\delta_{kk^{\prime}}\,\delta_{jj^{\prime}}\,\delta(t-t')$.
The last equal sign in \prettyref{eq:ExtInputToNeuron} follows from
the fact that two Gaussian white noise processes are equal if they
have equal first and second moments. The population averaged input
of the $i$th population is given by 
\begin{align}
y_{\mathrm{ext},i}(t) & =\frac{1}{M_{i}}\sum_{j=1}^{M_{i}}\left(N_{\mathrm{ext},i}\nu_{\mathrm{ext}}+\sqrt{N_{\mathrm{ext},i}\nu_{\mathrm{ext}}}\chi_{\mathrm{ext},i}^{j}(t)\right)\nonumber \\
 & =N_{\mathrm{ext},i}\nu_{\mathrm{ext}}+\frac{1}{M_{i}}\sum_{j=1}^{M_{i}}\sqrt{N_{\mathrm{ext},i}\nu_{\mathrm{ext}}}\chi_{\mathrm{ext},i}^{j}(t)\nonumber \\
 & =N_{\mathrm{ext},i}\nu_{\mathrm{ext}}+\sqrt{\frac{N_{\mathrm{ext},i}\nu_{\mathrm{ext}}}{M_{i}}}\chi_{\mathrm{ext},i}(t),\label{eq:ExtInputToPopulation}
\end{align}
where $M_{i}$ is the number of neurons in population $i$. The statistics
of the input therefore does not depend on whether each population
receives input form several Poisson sources with low firing rates
or one Poisson source with high firing rate. 

When the firing rate of the external input is modulated in time $\nu_{\mathrm{ext}}(t)=\nu_{\mathrm{ext}}f(t)$,
a neuron in population $i$ receives Gaussian white noise with time
dependent mean and variance 
\begin{equation}
y_{\mathrm{ext},i}(t)=\mu_{\mathrm{ext},i}(t)+\sigma_{\mathrm{ext},i}(t)\chi_{\mathrm{ext},i}(t),\label{eq:ExtInputTimeDep}
\end{equation}
with $\mu_{\mathrm{ext},i}(t)=N_{\mathrm{ext},i}\nu_{\mathrm{ext}}(t)$
and $\sigma_{\mathrm{ext},i}(t)=\sqrt{N_{\mathrm{ext},i}\nu_{\mathrm{ext}}(t)/M_{i}}$.
The description introduced here is valid around one working point
of the populations and therefore for fluctuations around one stationary
firing rate. Motivated by this, we choose $f(t)$ to describe a modulation
around the external rate, which does not change its original value
when averaged over long time series, i.e. $\underset{T\rightarrow\infty}{\mathrm{lim}}\frac{1}{2T}\int_{-T}^{T}\nu_{\mathrm{ext}}(t)dt=\nu_{\mathrm{ext}}$.

Incorporating the external input into the self-consistency of the
rate fluctuations (\prettyref{eq:rate_convolution}) yields

\begin{equation}
r_{i}(t)=h_{i}\ast\sum_{j=1}^{N}M_{ij}^{A}y_{j}(\circ-d_{ij})+\tau_{m}J_{\mathrm{ext}}\left[h_{i}\ast y_{\mathrm{ext},i}(\circ)\right](t),\label{eq:rate_convolution_with_ext}
\end{equation}
with the strength of the external current $w_{ext}$ ($J_{\mathrm{ext}}=\taus w_{\mathrm{ext}}/C_{m}$)
in Ampere, which is chosen to equal the synaptic strength within the
network as defined in \prettyref{eq:diffeq_iaf}. Defining the effective
connectivity matrix $\tilde{\mathbf{M}}_{\mathrm{d}}(\omega)$ to
incorporate the convolution of the dynamic transfer function $h_{i}$
and the anatomical connectivity $M_{ij}^{\mathrm{A}},$ as well as
the delays and delay distribution \citep{Bos16_1}, the fluctuating
rate of the circuit reads in Fourier space 
\begin{align}
 & \mathbf{Y}(\omega)=\tilde{\mathbf{M}}_{\mathrm{d}}(\omega)\mathbf{\mathbf{Y}}(\omega)+\mathbf{D}_{\mathrm{ext}}(\omega)\mathbf{Y}_{\mathrm{ext}}(\omega)+\mathbf{X}(\omega)\nonumber \\
\Leftrightarrow & \mathbf{Y}(\omega)=\mathbf{P}(\omega)\left(\mathbf{D}_{\mathrm{ext}}(\omega)\mathbf{Y}_{\mathrm{ext}}(\omega)+\mathbf{X}(\omega)\right),\label{eq:mu_sigma_conv_fourier}
\end{align}
where the eigenvalues and eigenvectors of $\mathbf{P}(\omega)=(\mathbb{I}-\tilde{\mathbf{M}}_{\mathrm{d}}(\omega))^{-1}$
are defined in \prettyref{eq:decomp_propagator} and $\mathbf{D}_{\mathrm{ext}}(\omega)$
is a diagonal matrix with elements $D_{\mathrm{ext},ii}(\omega)=\tau_{m}J_{\mathrm{ext}}H_{i}(\omega)$.
The population rate spectra read
\begin{align}
\mathbf{C}_{\mathrm{I}}(\omega) & =\langle\mathbf{Y}(\omega)\mathbf{Y}^{T}(-\omega)\rangle\nonumber \\
 & =\underset{=\mathbf{C}_{0}(\omega)}{\underbrace{\mathbf{P}(\omega)\langle\mathbf{X}(\omega)\mathbf{X}^{T}(-\omega)\rangle\mathbf{P}^{T}(-\omega)}}+\underset{=\mathbf{C}_{\mathrm{ext}}(\omega)}{\underbrace{\mathbf{P}(\omega)\mathbf{D}_{\mathrm{ext}}(\omega)\langle\mathbf{Y}_{\mathrm{ext}}(\omega)\mathbf{Y}_{\mathrm{ext}}^{T}(-\omega)\rangle\mathbf{D}_{\mathrm{ext}}^{T}(-\omega)\mathbf{P}^{T}(-\omega)}},\label{eq:spec_network}
\end{align}
where we used that the internal and external noise sources have zero
mean and we assumed them to be uncorrelated. The spectrum observed
in the circuit is thus given by the spectrum generated within the
circuit (\prettyref{eq:spectrum}) and the spectrum imposed from the
outside $\mathbf{C}_{\mathrm{ext}}(\omega)$. Before calculating the
expectation value of the external fluctuations $\langle\mathbf{Y}_{\mathrm{ext}}(\omega)\mathbf{Y}_{\mathrm{ext}}^{T}(-\omega)\rangle$,
we notice that the autocorrelation of the signal defined in \prettyref{eq:ExtInputTimeDep}
depends on two time arguments, namely the time-lag $\tau$ as well
as the global time $t$, which is induced by the modulation of the
rate
\begin{align}
\langle y_{\mathrm{ext},i}(t)y_{\mathrm{ext},j}(t+\tau)\rangle & =N_{\mathrm{ext},i}N_{\mathrm{ext},j}\nu_{\mathrm{ext}}^{2}f_{i}(t)f_{j}(t+\tau)+\delta_{ij}\delta(\tau)\,\frac{N_{\mathrm{ext},i}\nu_{\mathrm{ext}}}{M_{i}}f_{i}(t),\label{eq:autocorr_ext}
\end{align}
where we used that $\langle\chi_{\mathrm{ext},i}(t)\chi_{\mathrm{ext},j}(t+\tau)\rangle=\delta_{ij}\delta(\tau)$.
The autocorrelation is modulated in time and the process is therefore
not stationary. The time dependent spectral density, also known as
the Wigner-Ville spectrum (GWVS) \citep{Matz03} is given by

\begin{align}
\langle\mathbf{Y}_{\mathrm{ext}}(\omega,t)\mathbf{Y}_{\mathrm{ext}}^{T}(-\omega,t)\rangle_{ij} & =\int_{-\infty}^{\infty}e^{-i\omega\tau}\langle y_{\mathrm{ext},i}(t)y_{\mathrm{ext},j}(t+\tau)\rangle d\tau\nonumber \\
 & \overset{t'=t+\tau}{=}N_{\mathrm{ext},i}N_{\mathrm{ext},j}\nu_{\mathrm{ext}}^{2}f_{i}(t)e^{i\omega t}\underset{T\rightarrow\infty}{\mathrm{lim}}\int_{-T}^{T}e^{-i\omega t'}f_{j}(t')\,dt'\nonumber \\
 & \quad\quad\:+\delta_{ij}\delta(\tau)\,\frac{N_{\mathrm{ext},i}\nu_{\mathrm{ext}}}{M_{i}}\,f_{i}(t)\nonumber \\
 & =N_{\mathrm{ext},i}N_{\mathrm{ext},j}\nu_{\mathrm{ext}}^{2}f_{i}(t)e^{i\omega t}F_{j}(\omega)+\delta_{ij}\delta(\tau)\frac{N_{\mathrm{ext},i}\nu_{\mathrm{ext}}}{M_{i}}f_{i}(t),\label{eq:spec_t}
\end{align}
where $F(\omega)$ denotes the Foureir transform of $f(t)$. The GWVS
describes the time-frequency distribution of the mean energy of the
signal $y_{\mathrm{ext}}(t)$. The normalized marginal distribution
can be obtained by taking the average over time

\begin{align}
\langle\mathbf{Y}_{\mathrm{ext}}(\omega)\mathbf{Y}_{\mathrm{ext}}^{T}(-\omega)\rangle_{ij} & =\underset{T\rightarrow\infty}{\mathrm{lim}}\frac{1}{2T}\int_{-T}^{T}\langle\mathbf{Y}_{\mathrm{ext}}(\omega,t)\mathbf{Y}_{\mathrm{ext}}^{T}(-\omega,t)\rangle_{ij}dt\nonumber \\
 & =\mu^{2}F_{i}(\omega)\underset{T\rightarrow\infty}{\mathrm{lim}}\frac{1}{2T}\int_{-T}^{T}f_{j}(t)e^{i\omega t}dt+\sigma^{2}\underset{T\rightarrow\infty}{\mathrm{lim}}\frac{1}{2T}\int_{-T}^{T}f_{i}(t)dt\nonumber \\
 & =N_{\mathrm{ext},i}N_{\mathrm{ext},j}\nu_{\mathrm{ext}}^{2}\underset{T\rightarrow\infty}{\mathrm{lim}}\frac{1}{2T}F_{i}^{*}(\omega)F_{j}(\omega)\nonumber \\
 & \quad+\delta_{ij}\frac{N_{\mathrm{ext},i}\nu_{\mathrm{ext}}}{M_{i}}\underset{T\rightarrow\infty}{\mathrm{lim}}\frac{1}{2T}\int_{-T}^{T}f_{i}(t)dt\nonumber \\
 & =N_{\mathrm{ext},i}N_{\mathrm{ext},j}\nu_{\mathrm{ext}}^{2}\underset{T\rightarrow\infty}{\mathrm{lim}}\frac{1}{2T}F_{i}^{*}(\omega)F_{j}(\omega)+\delta_{ij}\frac{N_{\mathrm{ext},i}\nu_{\mathrm{ext}}}{M_{i}},\label{eq:spec_marg}
\end{align}
where the last integral could be discarded since we required that
the time dependent modulation of the firing rate $f_{i}(t)$ averages
out over long time series. The terms above show that a modulation
of the external firing rates yields two contributions to the spectrum.
The first term describes the contribution that arises from the modulation
of the mean, which can give rise to new peaks. In the original microcircuit
model, the external rate is constant ($f_{i}(t)=1$). The first term
thus solely gives a contribution at zero frequency ($F_{i}(\omega)=2\pi\delta(\omega)$).
The second term describes the effect of the modulation of the variance.
This term cannot introduce new peaks in the spectrum, since it does
not depends on $\omega$. It, however, gives a contribution at all
frequencies and can therefore influence the amplification of the dynamical
modes of the systems. The contribution of the external variance is
large compared to the variance of the noise generated within the network,
which is of order $\nu_{\mathrm{ext}}/M_{i}$. However, its contribution
is negligible, since it is small for in the microcircuit model and
additionally filtered out by the transfer function of the population
for larger frequencies (see definition of $\mathbf{D}_{\mathrm{ext}}(\omega)$
and \prettyref{eq:spec_network}). In the following we will therefore
focus on the dynamical contribution of the first term.

Let us now assume that the fraction $a$ of the external input to
the $k$th population is modulated by a sinusoid of frequency $\omega_{\I}$
such that $f(t)=1+a\:\sin(\omega_{\I}t)$ yielding the modulated mean
and variance

\begin{align}
\mu_{\mathrm{ext},k}(t) & =N_{\mathrm{ext},k}\nu_{\mathrm{ext}}\left(1+a\,\sin(\omega_{\I}t)\right)\nonumber \\
\sigma_{\mathrm{ext},k}^{2}(t) & =\frac{N_{\mathrm{ext},k}}{M_{k}}\nu_{\mathrm{ext}}\left(1+a\,\sin(\omega_{\I}t)\right),\label{eq:mu_var_mod_sinus}
\end{align}
while all other populations receive unmodulated external input. The
contribution of the mean modulation to the spectrum of the external
signal is then, for $\omega>0$, given by

\begin{equation}
\langle\mathbf{Y}_{\mathrm{ext}}(\omega)\mathbf{Y}_{\mathrm{ext}}^{T}(-\omega)\rangle_{ij}=\begin{cases}
N_{\mathrm{ext},k}^{2}\nu_{\mathrm{ext}}^{2}\underset{T\rightarrow\infty}{\mathrm{lim}}\frac{1}{2T}a^{2}\pi^{2}\left(\delta^{2}(\omega-\omega_{\I})+\delta^{2}(\omega+\omega_{\I})\right) & \mbox{for}\:i=j=k\\
0 & \mbox{else.}
\end{cases}\label{eq:spec_ext_inf_time}
\end{equation}

Thus the spectrum exhibits a $\delta$-peak at the frequency of the
modulating signal. To determine the height of the peak, we consider
that the measurement or the modulation lasts only a finite duration.
Inserting the definition of the $\delta$-function $2\pi\delta(\omega-\omega_{\I})=\underset{T\rightarrow\infty}{\mathrm{lim}}\int_{-T}^{T}e^{-i(\omega-\omega_{\I})\tau}d\tau$,
the peak height at $\omega_{\I}$, when measured in the interval $[0,T]$
(which changes the normalization constant in \prettyref{eq:spec_ext_inf_time}
from $1/2T$ to $1/T$), is given by

\begin{equation}
\underset{\omega\rightarrow\omega_{\I}}{\mathrm{lim}}\delta_{T}(\omega-\omega_{\I})=\frac{1}{2\pi}\underset{\omega\rightarrow\omega_{\I}}{\mathrm{lim}}\int_{0}^{T}e^{-i(\omega-\omega_{\I})\tau}d\tau=\frac{1}{2\pi}\underset{\omega\rightarrow\omega_{\I}}{\mathrm{lim}}\frac{e^{-i(\omega-\omega_{\I})T}-1}{-i(\omega-\omega_{\I})}=\frac{T}{2\pi},\label{eq:finite_delta}
\end{equation}
yielding the following spectrum of the external signal

\begin{equation}
\langle\mathbf{Y}_{\mathrm{ext}}(\omega)\mathbf{Y}_{\mathrm{ext}}^{T}(-\omega)\rangle_{ij}=\begin{cases}
N_{\mathrm{ext},k}^{2}\nu_{\mathrm{ext}}^{2}a^{2}\frac{T}{4} & \mbox{for}\:i=j=k\:\mbox{and }\:\omega=\omega_{\I}\\
0 & \mbox{else.}
\end{cases}\label{eq:spec_ext_finite_time}
\end{equation}
The additional contribution to the spectrum visible in the network
due to the external input is therefore given by
\begin{align}
C_{\mathrm{ext},ij}(\omega) & =\sum_{n,m}P_{in}(\omega)D_{\mathrm{ext},nn}(\omega)\langle Y_{\mathrm{ext}}(\omega)Y_{\mathrm{ext}}^{T}(-\omega)\rangle_{nm}D_{\mathrm{ext},mm}(-\omega)P_{jm}(-\omega)\label{eq:spec_ext_elements}
\end{align}
yielding 
\begin{equation}
\mathbf{C}_{\mathrm{ext}}(\omega)=\begin{cases}
w_{\mathrm{ext}}^{2}N_{\mathrm{ext},k}^{2}\nu_{\mathrm{ext}}^{2}a^{2}\frac{T}{4}\left|H_{k}(\omega)\right|^{2}\mathbf{P}(\omega)\mathbf{P}^{T}(-\omega) & \mbox{for}\:\omega=\omega_{\I}\\
0 & \mbox{else.}
\end{cases}\label{eq:spec_ext_final}
\end{equation}
The expression above shows that the peak imposed on one population
from the outside is propagated through the network and is therefore
visible in all populations. The height of the peaks scales linearly
with time.

As opposed to the above described current modulation, rate modulating
input is directly added on top of the population rate and not filtered
by the transfer function of the population, yielding the following
additive term in the spectrum
\begin{equation}
\mathbf{C}_{\mathrm{ext}}^{\mathrm{RM}}(\omega)=\begin{cases}
w_{\mathrm{ext}}^{2}N_{\mathrm{ext},k}^{2}\nu_{\mathrm{ext}}^{2}a^{2}\frac{T}{4}\mathbf{P}(\omega)\mathbf{P}^{T}(-\omega) & \mbox{for}\:\omega=\omega_{\I}\\
0 & \mbox{else.}
\end{cases}\label{eq:spec_ext_rm_final}
\end{equation}
Analyzing the current modulating input further by decomposing the
externally imposed spectrum, when population $k$ is stimulated in
the eigenbasis of the propagator matrix, as already done for the internally
generated spectrum \prettyref{eq:spectrum_decomp}, yields

\begin{align}
\mathbf{D}_{\mathrm{ext}}(\omega)\langle\mathbf{Y}_{\mathrm{ext}}(\omega)\mathbf{Y}_{\mathrm{ext}}^{T}(-\omega)\rangle\mathbf{D}_{\mathrm{ext}}(-\omega) & =\sum_{n,km}\beta_{\mathrm{ext},nm}(\omega)\mathbf{u}_{n}(\omega)\mathbf{u}_{m}^{T*}(\omega)\label{eq:Dext_decomp}
\end{align}

with 
\begin{equation}
\beta_{\mathrm{ext},nm}(\omega)=w_{\mathrm{ext}}^{2}N_{\mathrm{ext},k}^{2}\nu_{\mathrm{ext}}^{2}a^{2}\frac{T}{4}\left|H_{k}(\omega)\right|^{2}\alpha_{n}^{k}(\omega)\alpha_{m}^{k*}(\omega)\label{eq:beta_ext}
\end{equation}

and $\alpha_{j}^{k}(\omega)=\mathbf{v}_{j}^{\mathrm{T}}(\omega)\mathbf{e}_{k}$.
Inserting this into \prettyref{eq:spec_ext_final} yields the decomposition
of the population rate spectra at $\omega_{\I}$

\begin{align}
\mathbf{C}_{\mathrm{I}}(\omega) & =\sum_{n,m}\frac{\beta_{nm}(\omega)+\beta_{\mathrm{ext},nm}(\omega)}{(1-\lambda_{n}(\omega))(1-\lambda_{m}^{*}(\omega))}\mathbf{u}_{n}(\omega)\mathbf{u}_{m}^{\mathrm{T}*}(\omega).\label{eq:spectrum_decomp_with_input}
\end{align}
The expression above is referred to as the response spectrum. The
excess spectrum is defined as the additional power due to the input
$\delta\mathbf{C}(\omega)=\mathbf{C}_{\mathrm{I}}(\omega)-\mathbf{C}_{0}(\omega)$.
The power ratio, which describes the response spectrum at stimulus
frequency $\omega_{\I}$ normalized by the original spectrum at that
frequency, is evaluated at $\omega=\omega_{\I}$ and given by

\begin{equation}
\rho(\omega)=\frac{\mathbf{C}_{\mathrm{I}}(\omega)}{\mathbf{C}_{0}(\omega)}=1+\frac{\delta\mathbf{C}(\omega)}{\mathbf{C}_{0}(\omega)}.\label{eq:power_ratio}
\end{equation}

\subsection*{Approximation of the LFP\label{subsec:LFP}}

The LFP is described as the input to pyramidal cells. The rate fluctuations
received by pyramidal cells are given by

\begin{align*}
Y_{\mathrm{AMPA}}(\omega) & =W_{\E\E}H_{\mathrm{AMPA}}(\omega)Y_{\E}(\omega)=\frac{ae^{-i\omega d}}{1+i\omega\tau_{\mathrm{AMPA}}}Y_{\E}(\omega)\\
Y_{\mathrm{GABA}}(\omega) & =W_{\E\I}H_{\mathrm{GABA}}(\omega)Y_{\I}(\omega)=\frac{-be^{-i\omega d}}{1+i\omega\tau_{\mathrm{GABA}}}Y_{\I}(\omega),
\end{align*}
where $\mathbf{Y}(\omega)=(Y_{\mathrm{E}}(\omega),Y_{\mathrm{I}}(\omega))$
denotes the vector of fluctuating rates of the excitatory and the
inhibitory population. Here we assumed exponentially decaying synaptic
currents with time constants $\tau_{\mathrm{AMPA}}$ and $\tau_{\mathrm{GABA}}$.
The synaptic weights $a$ and $-b$ are defined in \prettyref{eq:conn_matrix-1}.
Mazzoni et al. \citep{Mazzoni2015} showed that the LFP is well approximated
by the sum of absolute values of the currents received by the pyramidal
neurons and is therefore given by

\begin{align}
C_{\mathrm{LFP}}(\omega) & =\langle\left(Y_{\mathrm{AMPA}}(\omega)-Y_{\mathrm{GABA}}(\omega)\right)\left(Y_{\mathrm{AMPA}}(-\omega)-Y_{\mathrm{GABA}}(-\omega)\right)\rangle\nonumber \\
 & =\frac{a^{2}}{1+\omega^{2}\tau_{\mathrm{AMPA}}^{2}}\langle\left|Y_{\E}(\omega)\right|^{2}\rangle+\frac{b^{2}}{1+\omega^{2}\tau_{\mathrm{GABA}}^{2}}\langle\left|Y_{\I}(\omega)\right|^{2}\rangle\nonumber \\
 & \quad+2ab\Re\left(\frac{\langle Y_{\E}(\omega)Y_{\I}(\omega)\rangle}{\left(1+i\omega\tau_{\mathrm{AMPA}}\right)\left(1-i\omega\tau_{\mathrm{GABA}}\right)}\right)\nonumber \\
 & =\underset{\coloneqq C_{\mathrm{LFP}}^{\mathrm{auto}}(\omega)}{\underbrace{\frac{a^{2}}{1+\omega^{2}\tau_{\mathrm{AMPA}}^{2}}C_{\E\E}(\omega)+\frac{b^{2}}{1+\omega^{2}\tau_{\mathrm{GABA}}^{2}}C_{\I\I}(\omega)}}\nonumber \\
 & \quad+\underset{\coloneqq C_{\mathrm{LFP}}^{\mathrm{cross}}(\omega)}{\underbrace{2ab\Re\left(\frac{C_{E\I}(\omega)}{\left(1+i\omega\tau_{\mathrm{AMPA}}\right)\left(1-i\omega\tau_{\mathrm{GABA}}\right)}\right)}}.\label{eq:def_CLFP_method}
\end{align}
The LFP is thus determined by the autocorrelation of the rate fluctuations
of each population as well as their crosscorrelation. The synaptic
dynamics induces an additional low-pass filtering of the contributions.
In this study we restrict the analysis of the LFP to $\delta$-synapses
($\tau_{\mathrm{AMPA}}=\tau_{\mathrm{GABA}}=0$) to isolate the phenomena
induced by the static network structure.

\section*{Acknowledgement}
The authors gratefully acknowledge the computing time granted (jinb33) by the JARA-HPC Vergabegremium and provided on the JARA-HPC Partition part of the supercomputer JUQUEEN at Forschungszentrum Jülich. Partly supported by Helmholtz Portfolio Supercomputing and Modeling for the Human Brain (SMHB), the Helmholtz young investigator group VH-NG-1028, EU Grant 269921 (BrainScaleS). This project received funding from the European Union's Horizon 2020 research and innovation programme under grant agreement No. 720270. All network simulations were carried out with NEST (http://www.nest-simulator.org).

\newpage{}

\end{document}